\documentclass[a4paper,11pt]{article}
\pdfoutput=1 
\usepackage{jcappub} 
\usepackage[T1]{fontenc} 
\usepackage{float}
\usepackage{hyperref}
\usepackage{comment}
\usepackage{tikz}
\usetikzlibrary{decorations.pathmorphing}
\title{Light Shining Through Wall Bounds on Axions From Obscured Magnetars
}

\usepackage{multirow}
\usepackage{multicol}

\usepackage{mathtools}

\def\ad{\Gamma}

\author[a]{Dibya S. Chattopadhyay,}
\author[a]{Basudeb Dasgupta,}
\author[a]{Amol Dighe}
\author[a,b]{and Mayank Narang}

\affiliation[a]{Tata Institute of Fundamental Research, \\Homi Bhabha Road, Colaba, Mumbai 400005, India}
\affiliation[b]{Academia Sinica Institute of Astronomy \& Astrophysics, \\11F of Astro-Math Bldg., No. 1, Sec. 4, Roosevelt Road, Taipei 10617, Taiwan, R.O.C.}

\emailAdd{d.s.chattopadhyay@theory.tifr.res.in}
\emailAdd{bdasgupta@theory.tifr.res.in}
\emailAdd{amol@theory.tifr.res.in}
\emailAdd{mnarang@asiaa.sinica.edu.tw}

\abstract{Coupling of axions or axion-like particles (ALPs) with photons may lead to photons escaping optically opaque regions by oscillating into ALPs. This phenomenon may be viewed as the Light Shining through Wall (LSW) scenario. While this LSW technique has been used previously in controlled laboratory settings to constrain the ALP-photon coupling ($g_{a\gamma}$), we show that this can also be applied in astrophysical environments. We find that obscured magnetars in particular are excellent candidates for this purpose. A fraction of photons emitted by the magnetar may convert to ALPs in the magnetar neighborhood, cross the large absorption column densities, and convert back into photons due to the interstellar magnetic field. Comparing the observed flux with the estimated intrinsic flux from the magnetar, we can constrain the contribution of this process, and hence constrain $g_{a\gamma}$. The effects of resonant conversion near the magnetar as well as ALP-photon oscillations in the interstellar medium are carefully considered. Taking a suitable magnetar candidate PSR J1622-4950, we find that the ALP-photon coupling can be constrained at $g_{a\gamma} \lesssim (10^{-10} - 10^{-11})$ GeV$^{-1}$ for low mass axions ($m_a \lesssim 10^{-12}$ eV). Our study reveals the previously unrealized potential for employing the LSW technique for obscured magnetars for probing and constraining ALP-photon couplings.
}

\keywords{axions; axion-like particles; light shining through wall bounds; axion constraints; magnetars} 

\begin{document}
\maketitle
\flushbottom


\section{\label{sec:Introduction}Introduction}

The quest to  understand the structures and interactions of elementary building blocks of Nature has led to many intriguing  extensions of the Standard Model (SM) of particle physics. Axions, first proposed as a solution to the strong CP problem in Quantum Chromodynamics (QCD)~\cite{PQ,Peccei:1977ur,Wilczek,Weinberg}, have been studied over the last four decades. The models of QCD axions, like the KSVZ~\cite{Kim:1979if,Shifman:1979if} and the DFSZ~\cite{Dine:1981rt,Zhitnitsky:1980tq} models were some of the first ones to be explored.
These axions have also been proposed as a viable dark matter candidates~\cite{Preskill:1982cy,Abbott:1982af,Dine:1982ah}.
A recent review on further theoretical constructions of the QCD axions can be found in~\cite{DiLuzio:2020wdo}.

The broader class of axion-like particles (ALPs), which are fundamental pseudoscalars that need not resolve the strong CP problem, has emerged as a prominent candidate for physics beyond the Standard Model (BSM).
These ALPs have been studied both within the dark matter paradigm~\cite{Duffy:2009ig,Arias:2012az,Chadha-Day:2021szb,Semertzidis:2021rxs,Adams:2022pbo} as well as outside of it~\cite{Marsh:2015xka,Choi:2020rgn}.
They are also predicted in the low energy limits of many string theory models~\cite{Svrcek:2006yi,Arvanitaki:2009fg,Acharya:2010zx,Ringwald:2012cu,Kamionkowski:2014zda,Stott:2017hvl,Halverson:2019cmy}.

Axions or ALPs in general are an additional pseudoscalar field. This pseudoscalar field, usually light and feebly interacting in nature, when coupled to photons, has a Lagrangian density~\cite{Raffelt:1987im,Raffelt:1996wa,kuster2007axions}:
\begin{equation}
	\mathcal{L}_{\text{ALP}} = \frac{1}{2}\, \partial^\mu a \, \partial_\mu a - \frac{1}{2} m_a^2 \, a^2 - \frac{1}{4} g_{a\gamma} a \, F_{\mu\nu} \tilde{F}^{\mu\nu}\;.
	\label{eq:lagrangian}
\end{equation}
Here, $``a''$ represents the ALP field, $m_a$ denotes its mass, and $g_{a\gamma}$ denotes the ALP-photon coupling strength. The interaction between ALPs and electromagnetic fields lead to possible detectable signals due to ALP-photon mixing, ALP decay etc. The interaction term in the ALP Lagrangian can be expressed as
\begin{equation}
	\mathcal{L}_{a\gamma} =  - \frac{1}{4} g_{a\gamma} a \, F_{\mu\nu} \tilde{F}^{\mu\nu} = g_{a\gamma} \, \vec{E} \cdot \vec{B}\, a \;.
\end{equation}
In the presence of a  magnetic field perpendicular to the direction of propagation, the polarization of photon parallel to the magnetic field directly couples to the ALPs.

\begin{figure}[t]
	\centering
\begin{tikzpicture}
	\draw[thick,decorate, decoration={snake}] (-2,0.25) -- (-.5,0.25) node[midway, above] {$\gamma$};
	
	\filldraw[gray,thick] (.85,-1) rectangle (1.15,1);
	\node at (1.6,-0.5) {Wall};
	\draw[thick,dashed] (-.5,0.25) -- (2.5,0.25) node[pos=0.2, above] {$a$};
	
	\draw[thick,decorate, decoration={snake}] (2.5,0.25) -- (4,0.25) node[midway, above] {$\gamma$};
	
	\draw[thick,decorate, decoration={snake}] (-.65,-0.75) -- (-.5,0.25) node[pos=0.4, left] {$\vec{B}_{\text{I}}$};
	\draw[thick,decorate, decoration={snake}] (2.5,-0.75) -- (2.5,0.25) node[pos=0.4, right] {$\vec{B}_{\text{II}}$};
	
	\node[thick,draw, circle, fill=white, minimum size=6pt] at (-0.5,-0.75) {};
	\node[thick,draw, circle, fill=white, minimum size=6pt] at (2.5,-0.75) {};
	\draw[thick] (-0.4,-0.6) -- (-0.6,-0.9);
	\draw[thick] (-0.6,-0.6) -- (-0.4,-0.9);
	\draw[thick] (2.4,-0.6) -- (2.6,-0.9);
	\draw[thick] (2.6,-0.6) -- (2.4,-0.9);

	\fill[gray] (-2.3,.25) circle (0.15);
 	\begin{scope}[shift={(-2.3, 0.25)}]
	\foreach \angle in {0, 30, ..., 330} {
		\draw[gray, -, line width=1.5pt] (0,0) -- (\angle:0.3);
	}
\end{scope}

\draw[color=white,thick] (-2.3,.25) circle (0.16);

\draw[thick] (4.05,-0.25) -- (4.05,.75);

\foreach \y in {-0.25,-0.15,...,0.85} {
	\draw[thin] (4.05,\y) -- ++(0.1,-0.1);}

\node at (-3.25, .25) {Source};
\node at (5, .25) {Detector};
\end{tikzpicture}

	\caption{Schematic diagram representing a light shining through a wall (LSW) experiment. The photons emitted by the source partially convert to axions ($a$) in the $\vec{B}_{\text{I}}$ field. The axions produced may travel onward through the wall, encounter the $\vec{B}_{\text{II}}$ field, and partially re-convert back to photons. The detector searches for signals of such photons.}
	\label{fig:LSW}
\end{figure}
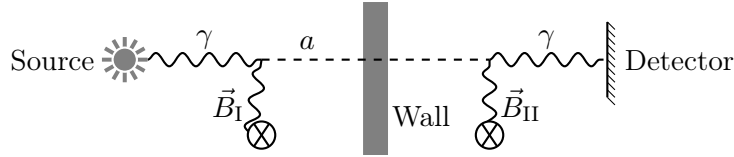

One of the first experimental search strategies for axions were outlined in~\cite{Sikivie:1983ip}. Extensive searches for detection of ALPs have been performed across a number of different settings~\cite{Raffelt:2006cw,Jaeckel:2010ni,Graham:2015ouw,Irastorza:2018dyq}, including  astrophysical observations and laboratory-based methods. Astrophysical observations of celestial objects such as white dwarfs, red giants, and active galactic nuclei provide key constraints on ALP coupling to different SM particles. Production of ALPs through mechanisms like the Primakoff process~\cite{Primakoff:1951iae} at such sources itself can be tightly constrained from arguments involving cooling and heat transport~\cite{RAFFELT19901,Friedland:2012hj,Giannotti:2017hny,Carenza:2020zil,DiLuzio:2021ysg,Dolan:2022kul}.
The phenomenon of ALP-photon oscillations in the presence of magnetic fields can also be exploited to obtain direct bounds on $g_{a\gamma}$, as such oscillations can affect the energy spectrum and observed emission profiles of these astrophysical sources~\cite{Mirizzi:2007hr,Hooper:2007bq,Hochmuth:2007hk,DeAngelis:2007wiw,DeAngelis:2007dqd,DeAngelis:2011id,Horns:2012kw,Wouters:2013hua,Schlederer:2015jwa,Berg:2016ese,Marsh:2017yvc,Mukherjee:2018oeb,Reynolds:2019uqt,Mukherjee:2019dsu,Calore:2020tjw,Calore:2021hhn,Reynes:2021bpe,Pant:2022ead}. Strong constraints on the ALP-photon coupling have also been obtained by using novel combinations of the physics of axion production at source and axion propagation through a medium~\cite{Payez:2014xsa,Meyer:2020vzy,Dessert:2020lil,Xiao:2020pra,Hoof:2022xbe}.
The possibility of conversion of background ALPs into photons by magnetars and neutron stars can provide additional constraints ~\cite{Lai:2006af,Pshirkov:2007st,Hook:2018iia,Huang:2018lxq,Safdi:2018oeu,Leroy:2019ghm,Battye:2019aco,Buschmann:2019pfp,Edwards:2019tzf,Foster:2020pgt,Darling:2020plz,Battye:2021xvt,Wang:2021hfb,Battye:2021yue,Witte:2022cjj,Noordhuis:2023wid,Tjemsland:2023vvc,Xue:2023ejt,Prabhu:2023cgb}. The photon to ALP conversion may also leave imprints on the observed photon spectra emitted by a magnetar or a neutron star~\cite{Bondarenko:2022ngb}.
A comprehensive collection of the current ALP constraints can be found in~\cite{AxionLimits} and the references therein.

Laboratory-based experiments employ a wide variety of detection schemes to explore the parameter space for ALPs. For example, resonant cavity experiments exploit the resonant conversion of ALP dark matter candidates into detectable photons within a precisely tuned cavity and scan through the axion parameter range looking for such ALP dark matter candidates~\cite{Irastorza:2018dyq,Semertzidis:2021rxs,Adams:2022pbo}. Other techniques involve the direct detection of axion-induced signals, such as the decay of ALP dark matter into two photons~\cite{Blout:2000uc,Boyarsky:2006ag,Grin:2006aw,Boyarsky:2007ay,Cadamuro:2011fd,Perez:2016tcq,Slatyer:2016qyl,Poulin:2016anj,Ng:2019gch,Roach:2019ctw,Thorpe-Morgan:2020rwc,Bolliet:2020ofj,Foster:2021ngm,Dekker:2021bos,Wadekar:2021qae,Keller:2021zbl,Balazs:2022tjl,Nakayama:2022jza,Roach:2022lgo,Bernal:2022xyi,Bessho:2022yyu,Langhoff:2022bij,Calore:2022pks,Carenza:2023qxh,Capozzi:2023xie,Todarello:2023hdk,Janish:2023kvi},
and the search for ALPs produced in the core of the Sun through Helioscopes~\cite{CAST:2004gzq,CAST:2017uph,IAXO:2020wwp}.

Another laboratory-based technique is the light-shining-through-wall (LSW) technique \mbox{\cite{VanBibber:1987rq,Hoogeveen:1990vq,Sikivie:2007qm,Redondo:2010dp}}, which does not depend upon the abundance of ALPs as DM candidates, but only on the ALP-photon coupling. This controlled and precise experimental approach employs a high-intensity laser beam that traverses a strong magnetic field in a specially designed setup. As shown in Fig.\,\ref{fig:LSW}, a fraction of the photons in the beam, travelling through the magnetic field ($\vec{B}_{\text{I}}$) gets converted into ALPs due to the Primakoff effect and is able to cross the optical barrier (as ALPs). On the other side of the barrier, due to the same conversion mechanism a fraction of the axions convert back into photons due to the magnetic field $\vec{B}_{\text{II}}$. If the resulting photons were to be captured by sensitive detectors, they would provide a telltale signal indicative of the presence of ALPs. The LSW technique offers a controlled laboratory environment to explore the parameter space for these elusive particles and has motivated a range of experimental efforts~\cite{Ehret:2010mh,Betz:2013dza,Ballou:2014myz,Ortiz:2020tgs,Dev:2021ofc}.

The LSW technique essentially boils down to looking for photons where there should be none. 
Note that the key ingredients necessary for an LSW experiment, as shown in Fig.\,\ref{fig:LSW}, are: (i) a luminous source, (ii) magnetic field between the source and the optical barrier, (iii) an opaque barrier, (iv) a second  magnetic field between the optical barrier and the detector, (v) a sensitive detector.

In this paper, we try to find suitable astrophysical environments where the LSW technique can be applied. We find that obscured magnetars~\cite{Turolla:2015mwa,Kaspi:2017fwg,Esposito:2018gvp} (i.e., magnetars whose luminosity is partially decreased due to a large absorption column density, corresponding to a large density of neutral hydrogen, in the foreground) can be a suitable candidate to constrain the ALP-photon coupling $g_{a\gamma}$.
 To constrain the ALP-photon coupling $g_{a\gamma}$, we consider how a non-zero ALP-photon coupling ($g_{a\gamma}$) can help photons escape thick layers of dust that otherwise block the light. We obtain the bound by imposing that the contribution from the $\gamma \to a \to \gamma$ process has to be less than the total observed flux of the photons,
\begin{equation}
	F (\gamma \to a \to \gamma) \lesssim F_{\text{obs}}   \quad \Rightarrow \quad P(\gamma \to a \to \gamma) \, F_{0} \lesssim F_{\text{obs}}  \;.
\end{equation}
Here, $F_{\text{obs}}$ is the observed flux, $F (\gamma \to a \to \gamma) \equiv P(\gamma \to a \to \gamma) \, F_{0}$ is the contribution from the $\gamma \to a \to \gamma$ process, and $F_{0}$ is the expected flux in the absence of any absorption, which can be estimated from the intrinsic luminosity estimates and the distance of the source from the observer. For the rest of this work, we denote $F_{0}$ as the ``intrinsic'' flux.

To estimate $P(\gamma \to a \to \gamma)$ we need to know the conversion probabilities before and after the optically opaque medium, which in this case is the nebula surrounding the magnetar.
We estimate the conversion probability $P(\gamma \to a)$ near the magnetar, with detailed discussion focusing upon the conditions and effects of resonant conversion. We also calculate the interstellar medium (ISM) conversion probability $P(a \to \gamma)$.
In spite of the large uncertainties in the survival probability and the magnetar neighborhood environment, we are able to obtain bounds on $g_{a \gamma} \lesssim (10^{-10} - 10^{-11}) \text{ GeV}^{-1}$, even in the conservative limit.
A better understanding of the primary sources of uncertainties can only lead to a more robust and possibly more efficient bound on  $g_{a \gamma}$.

The flow of this paper is as follows.
In section~\ref{sec:idea}, we discuss the suitability of the magnetars for the LSW technique implementation. In section~\ref{sec:magnetar}, we list the magnetar properties and estimate photon survival probability ranges, taking into account the uncertainties in modeling of the magnetar behavior.
In section~\ref{sec:theory}, we set up the formalism for calculating the ALP-photon conversion probabilities in the presence of multiple resonances in the magnetar neighborhood, and oscillation in the interstellar medium.
In section~\ref{sec:probabilities}, we calculate the ALP-photon conversion probabilities in the ISM and in the magnetar neighborhood, investigating in detail the dependence of the resonant conversion probability on electron number density and mass of ALPs. Furthermore, we also explore the effects of uncertainties on the conversion probabilities.
In section~\ref{sec:result}, we plot the constraints obtained in the $(m_a - g_{a\gamma})$ plane and compare it against the leading constraints, as well as constraints obtained through the lab based LSW experiments.
In section~\ref{sec:conclusion} we present a brief overview of our findings and conclude with how the constraint can be improved further.

\section{\label{sec:idea}Constraining $g_{a\gamma}$ from Obscured Magnetars}

In this section, we discuss the key similarities and differences between an LSW experiment in the controlled laboratory environment~\cite{Ehret:2010mh,Betz:2013dza,Ballou:2014myz,Ortiz:2020tgs,Dev:2021ofc} and an astrophysical LSW scenario that employs obscured magnetars in space. First, we list the key features that makes obscured magnetars suitable for the LSW approach.
\begin{table}[t]
	\centering
	\renewcommand{\arraystretch}{1.25}
	\begin{tabular}{|c|c|c|c|c|}
		\hline
		Diagram &Key ingredients & Laboratory & Obscured magnetars & Similarity \\
		\hline
		\multirow{5}{*}{
	\begin{tikzpicture}
	
		\foreach \angle in {0, 30, ..., 330} {
			\draw[gray, -,line width=1.5pt] (0,0) -- (\angle:0.3);
		}			
		\foreach \angle in {-15, 15, ..., 315} {
		\draw[white, -,line width=0.5pt] (0,0) -- (\angle:0.3);
	}		
		\fill[gray] (0,0) circle (0.15);
		\fill[gray] (0,0) circle (0.14);
		
		\draw[color=white,thick] (0,0) circle (0.16);

		\draw[thin,decorate, decoration={snake},->] (0,-.32) -- (0,-.85) node[pos=0.4, left] {$\gamma$};
		\draw[thin,dashed,->] (0,-.87) -- (0,-1.6) node[pos=0.4, left] {$a$};
		\draw[thin,decorate, decoration={snake},->] (0,-1.62) -- (0,-2.35) node[pos=0.6, left] {$\gamma$};
		\filldraw[gray, thin, opacity=0.5] (-.5,-.95) rectangle (.5,-1.32);
	 
		\draw[thin] (-0.3,-2.4) -- (0.32,-2.4);
	 
		 \foreach \x in {-0.25,-0.20,...,0.35} {
	 	\draw[thin] (\x,-2.4) -- ++(-0.1,-0.15);}
 	
   		\node at (0.45, -.7) {$\vec{B}_{\text{I}}$};
   		\node at (0.45, -1.7) {$\vec{B}_{\text{II}}$};
	\end{tikzpicture}
	}&Source   &    Laser      &    Magnetar     &   $\checkmark^{1}$      \\
		\cline{2-5}
		&Magnetic field  I  &       $\sim 10^{5} \text{ G}$   &      $\sim 10^{14} \text{ G}$     &     $\checkmark \checkmark$      \\
		\cline{2-5}
		&Opaque barrier    &     Wall (opaque)     &     Nebula (semi-opaque)&    $ \checkmark ^2$ \\
		\cline{2-5}
		&Magnetic field  II  &         $\sim 10^{5} \text{ G}$   &      $\sim 10^{-6} \text{ G}$       &    $\checkmark^3 $       \\
		\cline{2-5}
		&Detector    &    sensitive in-lab   &     X-ray observations     &      $\checkmark \checkmark$     \\
		\hline
	\end{tabular}
	\\ \vspace{5pt}
	\raggedright
	\footnotesize{$^1$ Large uncertainty in luminosity estimates of magnetars.}\\
	\footnotesize{$^2$ It is possible for some photons to always reach the detector due to the semi-opaque nature of the nebula.}\\
	\footnotesize{$^3$ The magnetic field in the ISM is weak ($\sim \mu$G), although the overall effect would be amplified due to large distances ($\sim$kpc) traversed.}\\
	\caption{A comparison between laboratory and magnetar environments to assess the applicability of the light shining through wall (LSW) technique in an astrophysical setting.}
	\label{table:comparison}
\end{table}

\subsection{Light Shining Through Wall in Astrophysics and Obscured Magnetars}

In Table.~\ref{table:comparison}, we compare the two different approaches, with $(\, \checkmark \,)$ denoting matching optimal conditions, and comments pointing out the presence of important caveats. Below, we further detail the comparisons made in Table~\ref{table:comparison}.
\begin{itemize}
	\item \textbf{Source}: The magnetar is a very luminous, albeit a less well understood source \mbox{\cite{Turolla:2015mwa,Kaspi:2017fwg,Esposito:2018gvp}}. 
	Note that an overestimation of the intrinsic source luminosity would lead to an overestimation of the intrinsic flux $F_0$ that would have been detected in the absence of any absorption. A higher estimation of $F_{0}$ leads to a more stringent bound on the efficiency of the $\gamma \to a \to \gamma$ contribution, and thus a stronger constraint on how large $P(\gamma \to a \to \gamma) $ can be. Therefore, to obtain a conservative bound, we must not overestimate the intrinsic luminosity.
	
	\item  \textbf{Magnetic field I}: The magnetic field in the magnetar neighborhood between the magnetar and the semi-opaque medium is extremely high $(|\vec{B}_{\text{I}}| \sim 10^{14} \text{ G})$~\cite{Turolla:2015mwa,Kaspi:2017fwg,Esposito:2018gvp}. Thus making it an ideal region for even possible resonant conversion of photons into ALPs.
	
	\item \textbf{Opaque barrier}: The nebula with high absorption column density $(N_H)$ plays the role of the opaque optical barrier in this scenario~\cite{Olausen:2013bpa}. Furthermore, the observed flux being significantly less than the intrinsic flux (estimated flux in the absence of any absorption) also suggests the existence of a semi-opaque barrier.
	Note that this is not a perfect barrier, but a semi-opaque one. In a laboratory based LSW experiment,  observing a non-zero photon flux at the detector would be a signature of the existence of ALPs. However, in the astrophysical scenario, in the absence of a perfect barrier we can \emph{only constrain} $g_{a\gamma}$. This is because, even in the absence of ALPs, some photons may reach the detector. 
	
	\item \textbf{Magnetic field II}: Between the opaque barrier and detector, we expect a fraction of the ALPs to convert back into photons due to the ISM magnetic field. However, the ISM magnetic field is weak, $\sim \text{ few} \times 10^{-6} \text{ G} $~\cite{Jansson_2012}, which leads to a small mixing and therefore a small conversion probability.
	
	\item \textbf{Detector}: For detecting X-ray emissions from magnetars, we can use observed flux data from space-based X-ray observatories like Swift, Chandra and XMM-Newton~\cite{Weisskopf:2000tx,2003SPIE.4851...28G,XMM:2001haf,Struder:2001bh,Turner:2000jy,Burrows:2005gfa}.
\end{itemize}
With the caveats regarding (i)~source luminosity uncertainties and (ii) the semi-opaque nature of the Nebula noted and discussed, we can now move to the implementation of the LSW technique in obscured magnetars.

\subsection{Photon Survival Probability and Axion Bounds}
\label{sec:threeprocesses}

To implement the LSW technique for obscured magnetars, we need both the observed flux $(F_{\text{obs}} )$ and the estimated intrinsic flux $(F_0)$. This would enable us to put an upper limit to the probability associated with the $\gamma \to a \to \gamma$ process.
We define the survival probability $P_{\text{sur}}(E_{\text{bin}} )$ as the ratio of the observed flux in the energy bin $E_{\text{bin}} \equiv (E,E+\delta E)$ and the intrinsic flux in the same bin:
\begin{equation}
	P_{\text{sur}} (E_{\text{bin}} ) = \frac{ F_{\text{obs}} (E_{\text{bin}} ) }{ F_{0} (E_{\text{bin}} ) }\;.
	\label{eq:Psur}
\end{equation}
The total observed flux $F_{\text{obs}}  (E_{\text{bin}} ) $ will be due to the following contributing processes:
\begin{itemize}
	\setlength\itemsep{0pt}
	\item[A.] Survival of some photons in the energy range $ (E,E+\delta E)$ due to the semi-opaque nature of the neutral hydrogen column density.
	\item[B.] Down-scattering of higher energy photons  in the nebula --- with initial energies in the range $(E+\delta E , E_{\max})$ --- leading to extra flux injection in the $ (E,E+\delta E) $ bin.  We ignore any upscattering in this analysis.
	\item[C.] Escape of a certain fraction of photons through the $\gamma \to a \to \gamma$ mechanism.
\end{itemize}
Therefore, the observed flux must be larger than or equal to the contribution from the above mentioned three processes
\begin{equation}
	F_{\text{obs}} (E_{\text{bin}})  = F^{\, \text{Att}} (E_{\text{bin}}) + F^{\, \text{DS}} (E_{\text{bin}}) + F^{\, \text{LSW}} (E_{\text{bin}})\;.
\end{equation}
Here, $F^{\, \text{Att}} (E_{\text{bin}}) $ is the attenuated flux contribution from the process A, while $F^{\, \text{DS}} (E_{\text{bin}})$ is the down-scattered flux contribution from process B, and $F^{\, \text{LSW}} (E_{\text{bin}})$ is the flux that escaped through the LSW mechanism $(\gamma \to a \to \gamma)$, corresponding to process C. The contributions from the process A and B are discussed in more detail in appendix~\ref{sec:contribution}.
\begin{figure}[t]
	\centering
	\includegraphics[width=\textwidth]{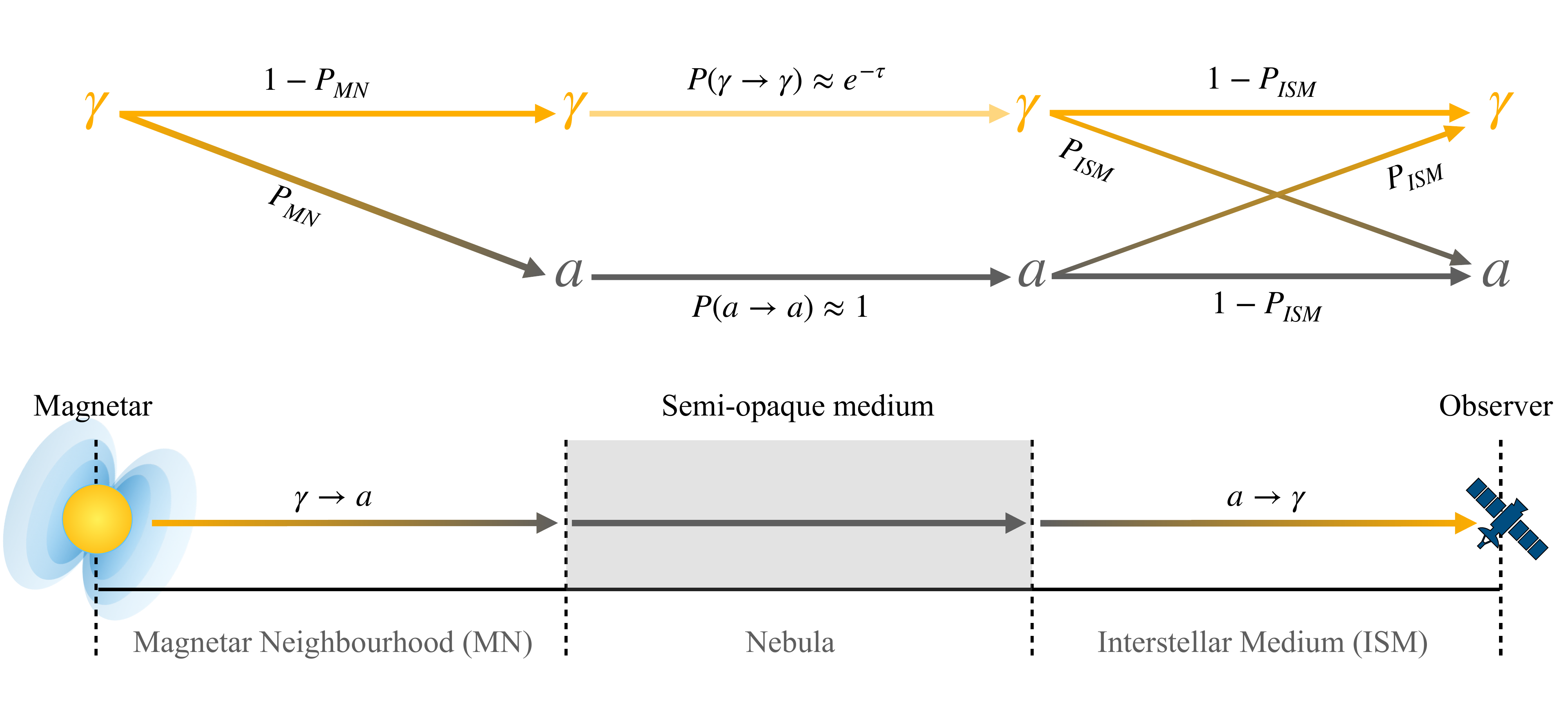}
	\caption{A simplified schematic diagram representing the different regions of obscured magnetar system (below) and the associated ALP-photon conversion probabilities (above). We illustrate all possible states of the system at different times during propagation. Here, $P_{MN}$ denotes the $\gamma \to a$ conversion probability in the magnetar neighborhood (MN), and $P_{ISM}$ denotes the $a\to \gamma$ conversion probability in the interstellar medium (ISM). The attenuation of photons in the nebula is represented by $e^{-\tau}$.}
	\label{fig:LSWastro}
\end{figure}
As both of these processes A and B have a non-negative contribution, we get
\begin{equation}
		F_{\text{obs}} (E_{\text{bin}})  \, \geq \, F^{\, \text{LSW}} (E_{\text{bin}})  \;.
\end{equation}
From Fig.\,\ref{fig:LSWastro}, we obtain (see appendix~\ref{sec:contribution})
\begin{equation}
	F^{\, \text{LSW}} (E_{\text{bin}}) \, \approx \, P_{MN} \cdot P_{ISM} \cdot F_{0} (E_{\text{bin}})\;,
\end{equation}
where $P_{MN} $ is the conversion probability in the magnetar neighborhood, and $P_{ISM}$ is the conversion probability in the ISM. Note that both of these probabilities would have an implicit energy dependence.
Here, we have ignored any effects of the conversion inside the Nebula, approximating the ALP survival probability as $P(a \to a) \approx 1$ (see appendix~\ref{sec:contribution} for justification). This leads to
\begin{equation}
	F_{\text{obs}} (E_{\text{bin}}) \, \geq  \, P_{MN} \cdot  P_{ISM} \cdot F_{0} (E_{\text{bin}})  \; .
\end{equation}
Since $P_{\text{sur}} (E_{\text{bin}})$ is defined as the ratio between  $F_{\text{obs}} (E_{\text{bin}})$ and $F_{0} (E_{\text{bin}}) $, we can write
\begin{equation}
	P_{\text{sur}} (E_{\text{bin}})  \, \gtrsim \,  P_{MN} \cdot P_{ISM}\; .
\end{equation}
For all scenarios where  $F_{\text{obs}}  (E_{\text{bin}}) \ll F_{0} (E_{\text{bin}})$, i.e. $P_{\text{sur}} (E_{\text{bin}}) \ll 1$, this would lead to a bound on the conversion probability between photons and ALPs, and therefore on the ALP-photon coupling $g_{a\gamma}$.

In section~\ref{sec:magnetar}, we determine $P_{\text{sur}} (E_{\text{bin}})$ using observations of a suitable magnetar candidate. In section~\ref{sec:theory}, we build up the formalism for the ALP-photon conversion probability in the presence of oscillations and possible resonances. In section~\ref{sec:probabilities}, we estimate the conversion probabilities $P_{MN}$ and $P_{ISM}$ as functions of physical parameters like the ALP-photon coupling $g_{a\gamma}$, ALP mass $m_a$, electron density, magnetic field strength and energy of the photon~$\omega_\gamma$. This then allows us to obtain the bound on $g_{a\gamma}$.

\section{The Magnetar Candidate:  PSR J1622-4950}
\label{sec:magnetar}
As discussed in the previous section, any suitable magnetar candidate where we can apply the LSW technique would need to be: (i) bright, (ii) highly obscured, i.e. with a small $P_{\text{sur}} (E_{\text{bin}})$, and (iii)~with significant magnetic field on both sides of the semi-opaque region. With such a magnetar, knowledge of its intrinsic luminosity and the measurement of its observed flux would allow us to calculate the photon survival probability $P_{\text{sur}}(E_{\text{bin}})$.

The magnetar PSR J1622-4950 satisfies all the criteria stated above, cf.~\cite{Anderson_2012,Olausen:2013bpa}: 
\begin{itemize}
	\item[(i)] The availability of estimates for the intrinsic flux $F_{0} (E_{\text{bin}})$, from modeling of the intrinsic total flux, the flux profile, and effects of absorption in the nebula.
	\item[(ii)] Large estimated absorption column density $N_H$ due to neutral hydrogen, which would lead to a highly absorbed signal in the soft X-ray range $\sim (0.1-1)$ keV. 
	\item[(iii)] High Magnetic field strength of $B_{\text{surface}} \sim 2.7 \times 10^{14}$ G near the magnetar, which would facilitate ALP-photon oscillations, in fact, as we will show in section~\ref{sec:probabilities}, the photons propagating outward in the magnetar neighborhood encounter resonances which further increases the conversions into ALPs.
	\item[(iv)] The availability of measurements of observed flux $F_{\text{obs}} (E_{\text{bin}})$ by XMM-Newton (Feb, 2011) in the soft X-ray regime $\sim (0.7-1)$ keV~\cite{XMM:2001haf,Struder:2001bh,Turner:2000jy}.
\end{itemize}
The basic properties of the magnetar PSR J1622-4950  are given in Table~\ref{table:magnetar}. More details about the system can be found in~\cite{Levin_2010,Anderson_2012,Camilo_2018,Chrimes:2022yik}.
We start by modeling the behavior of charge density and magnetic field in the magnetar neighborhood from the properties of Table~\ref{table:magnetar}. We then use the intrinsic flux estimate of the magnetar and the observed flux data, as given in~\cite{Anderson_2012}, to estimate the survival probability $P_{\text{sur}} (E_{\text{bin}})$.

\subsection{Charge Density and Magnetic Fields in the Magnetar Neighborhood}

\begin{table}[t]
	\centering
	\renewcommand{\arraystretch}{1.5} 
	\begin{tabular}{|c|c|c|c|c|}
		\hline
		Magnetar &$P $ (s) & $\dot{P}\; (10^{-11}  \text{ s s}^{-1})$ & $B_{\text{surface}} \; (10^{14} \text{ G})$ & $L $ (kpc)  \\
		\hline
		PSR J1622-4950 &$4.361(1)$ & $1.7(1)$ & $2.7$ & $\sim 9$ \\
		\hline
	\end{tabular}
	\caption{The period of rotation $P$, the spin-down time $\dot{P}$, the magnetic field strength $B_{\text{surface}}$ at the surface and the distance $L$ from Earth, for the magnetar PSR J1622-4950~\cite{Levin_2010,Anderson_2012,Olausen:2013bpa}.}
	\label{table:magnetar}
\end{table}

Given a magnetic field $\vec{B}$ in the neighborhood of a rotating magnetar, the minimum density required to sustain this magnetic field is the Goldreich-Julian (GJ) charge density~\cite{1969ApJ...157..869G,Thompson:2001ig}, given by
\begin{equation}
	n_{GJ} \approx \frac{\vec{\Omega} \cdot \vec{B}}{\sqrt{\pi \alpha}}\; .
\end{equation}
Here, $\alpha \approx 1/137$ is the fine structure constant, and $\Omega \equiv 2\pi/P$, where $P$ is the period of rotation of the magnetar. The GJ charge density at the surface of the magnetar PSR J1622-4950 is estimated to be
\begin{equation}
	n_{GJ} \approx \frac{\langle \vec{\Omega} \cdot \vec{B} \rangle }{\sqrt{\pi \alpha}} \approx 2.15 \times 10^{12} \text{ cm}^{-3}\;,
\end{equation}
where we have performed the averaging by using $|\vec{B}|_{\text{avg}} \approx B_{\text{surface}}/2 \approx 1.35 \times 10^{14}$ G.
However, this is only a lower bound on the charge density. Processes like pair production, plasma-instability and particle acceleration are expected to have a significant contribution to the total charge density near a magnetar.
From numerical simulations~\cite{Lyutikov:2007fn,Kalapotharakos_2012,Timokhin:2015dua,Cruz:2020vfm}, it has been estimated that the charge density at the surface can be higher than the GJ bound by a maximum factor of $\kappa \sim O(10^4-10^5)$. We therefore consider the range of charge density 
\begin{equation}
	n_0 \sim (10^{12} - 10^{18}) \text{ cm}^{-3}
\end{equation}
at the surface of the magnetar. 

\begin{table}[t]
	\centering
	\renewcommand{\arraystretch}{1.4}
	\begin{tabular}{|c|c|c|c|}\hline
		\multicolumn{4}{|c|}{PSR J1622-4950 (XMM, 2011-02-22) \qquad (``Absorbed BB'' model)} \\
		\hline
		Observed data & \multicolumn{2}{|c|}{Magnetar modeling} & Opaque region\\
		\hline
		$F_\text{obs}(\text{total})$ & $F_0(\text{total})$ & 	$kT$ & $N_H\,$ \\ 
		(erg cm$^{-2}$ s$^{-1}$) &(erg cm$^{-2}$ s$^{-1}$)& (keV) & ($10^{22} \,$cm$^{-2}$)  \\
		\hline
		$3.0^{+0.8}_{-0.6}\times 10^{-14}$ &$11^{+9}_{-4}\times 10^{-14}$& 	$0.5 \pm 0.1 $& $5.4^{+1.6}_{-1.4}$ \\
		\hline
	\end{tabular}
	\caption{Observed flux and estimated intrinsic flux from the magnetar PSR J1622-4950, with the absorbed blackbody (BB) modeling~\cite{Anderson_2012}. All the data given in the table are at 90\% confidence level. The absorption column density $N_H$ and the temperature $T$ is also a part of this model.  The ``total'' flux corresponds to the energy range 0.3-10.0 keV.}
	\label{table:flux}
\end{table}

The detailed behavior of the magnetic field near a magnetar is complex, and beyond the scope of our current work. However, the general behavior can be estimated by modeling this magnetic field as that of a dipole. We take the surface magnetic field strength to be $B_0 \equiv |\vec{B}|_{\text{avg}} \approx 1.35 \times 10^{14}$ G, at a surface radius $r_0 =10$ km, and model the spatial behavior of transverse component of the magnetic field as
\begin{equation}
	|\vec{B}_T (r)| \sim B_0 \left( \frac{r}{r_0} \right)^{-3}\; .
	\label{eq:Bspatial}
\end{equation}
The charge density, at its theoretical minimum, would have the same power law behavior as that of the magnetic field~\cite{1969ApJ...157..869G}. Even for the contributions from other processes, we expect the charge density to always fall faster than $r^{-2}$. We assume, for the sake of simplification, that the charge density $n_e$ falls as
\begin{equation}
	n_e (r) \sim n_0 \left( \frac{r}{r_0} \right)^{-3}\;.
	\label{eq:nspatial}
\end{equation}
Changing the power law behavior away from $r^{-3}$ by a small amount should not affect our results. This is because, as discussed in sections~\ref{sec:multres} and \ref{sec:pMagnetar}, the conversion probability in the magnetar neighborhood depends on the net effect of all the resonances and not on an individual resonance. The power law behavior assumed allows analytic estimation of the dependence of the conversion probability on $g_{a\gamma}$ in the magnetar neighborhood.

\subsection{Photon Survival Probability $P_{\text{sur}}$}

\begin{figure}[t]
	\centering
	\includegraphics[width=.5 \textwidth]{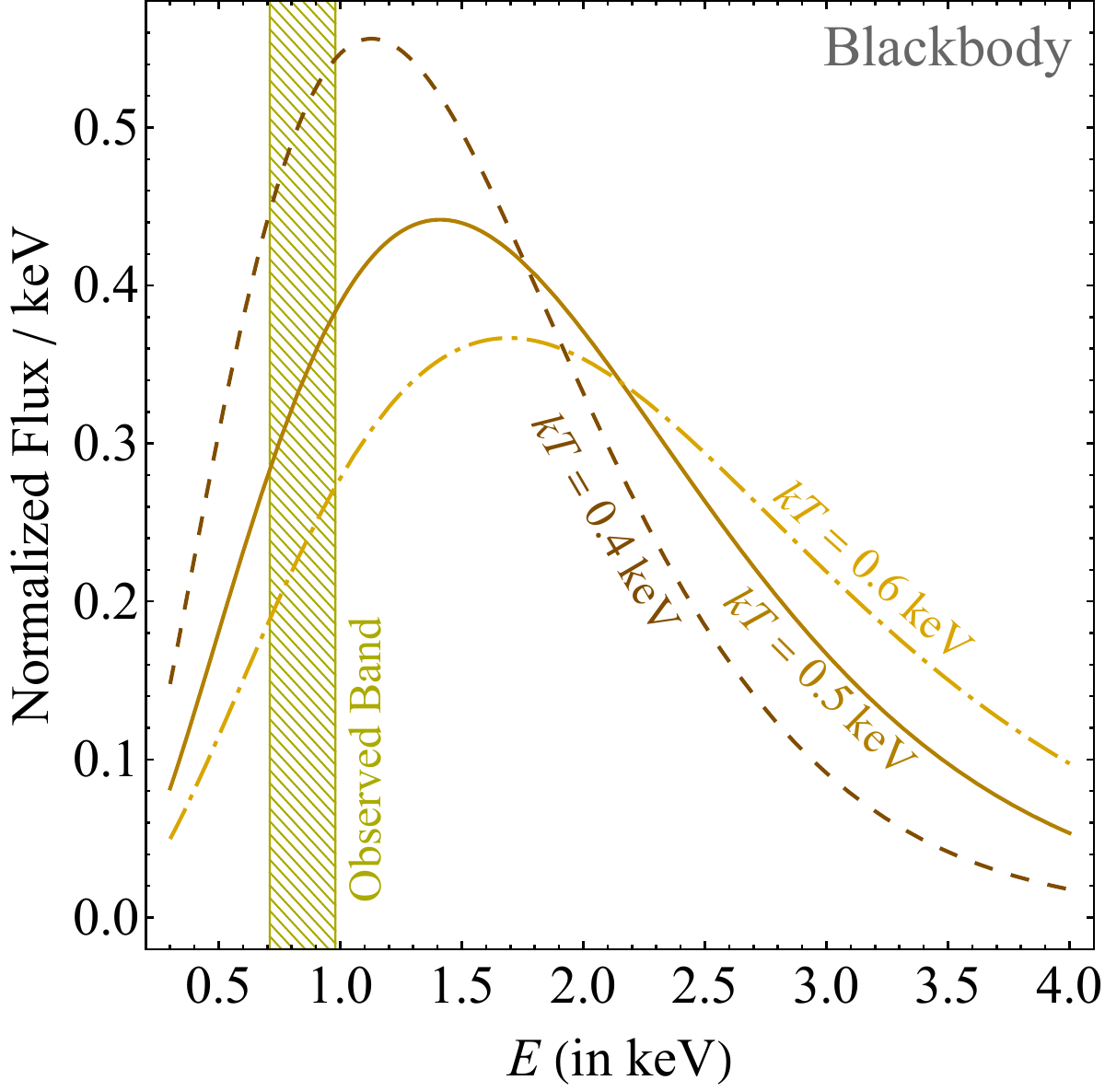}
	\caption{Determining the intrinsic flux in the observed bin using the absorbed blackbody (BB) modeling~\cite{Anderson_2012} of the obscured magnetar PSR J1622-4950.}
	\label{fig:BB}
\end{figure}

To determine the photon survival probability as defined in Eq.~(\ref{eq:Psur}), we use the measurement of the observed flux from the XMM-Newton 2011-02-22 observation of PSR J1622-4950~\cite{Anderson_2012}.
	While Table~\ref{table:flux} presents the total observed flux in the energy range of $(0.3 - 10)$ keV, for our analysis, we use the flux in the lowest energy bin of $(0.71 - 0.98)$ keV from Fig.\,1 of~\cite{Anderson_2012}. In this bin, the observed flux is lowest, which would lead to optimal results in our analysis. The flux in this bin is $(1.90-5.61) \times 10^{-9}$ photons~cm$^{-2}$~s$^{-1}$~keV$^{-1}$,  where the range corresponds to the error bars as shown in the figure. Taking $E_{\text{avg}} = 0.83$ keV, we find
\begin{equation}
	F_{\text{obs}} (E_\text{bin}) \approx (0.68 - 2.01) \times 10^{-18} \; \text{erg cm$^{-2}$ s$^{-1}$}\;.
\end{equation}

The total intrinsic flux given in Table~\ref{table:flux} is estimated in~\cite{Anderson_2012} by modeling the magnetar using a absorbed blackbody (BB) fit, where the observed flux is fit with a joint profile of blackbody radiation with temperature $T$ of the magnetar of the magnetar and the absorption column density $N_H$ in our line of sight as free parameters.
The fit parameters as given in Table~\ref{table:flux} allow us to reconstruct the spectrum of X-rays from the magnetar as shown in Fig.\,\ref{fig:BB}. From the figure, the intrinsic flux in the $(0.71 - 0.98)$ keV bin can be determined to be
\begin{equation}
	F_0 (E_\text{bin}) = (0.44 - 2.72) \times 10^{-14} \; \text{erg cm$^{-2}$ s$^{-1}$}\;,
\end{equation}
at $90\%$ confidence level.

From the values of $F_{\text{obs}} (E_\text{bin})$ and $F_0 (\text{bin}) $, we estimate the photon survival probability range to be
\begin{equation}
	P_{\text{sur}} (E_\text{bin}) = (0.25 - 4.58) \times 10^{-4} \;.
\end{equation}
The smallness of the photon survival probability\footnote{Note that these numbers have been obtained from an absorbed BB fit.
	The magnetar flux has also been modeled using the absorbed power law~\cite{Anderson_2012}, where the total intrinsic flux estimate and consequently the total survival probability in the bin has much larger uncertainties. Further, the maximum photon survival probability allowed by this model is significantly smaller than that obtained from the absorbed BB model. We choose to use the conservative bound obtained by the absorbed BB modeling.}
 would lead to strong constraints on any new physics process which would have an overall positive contribution to the number of photons that can escape the semi-opaque barrier.

In the next section we consider such a new physics scenario, where the $\gamma \to a \to \gamma$ oscillations induced by the ALP-photon coupling $g_{a\gamma}$ would contribute to the net photon survival probability.

\section{\label{sec:theory}Formalism for Photon-ALP conversion}
From the Lagrangian in Eq.~(\ref{eq:lagrangian}), denoting the direction of propagation as the z-direction, in the ultra-relativistic limit we obtain~\cite{Raffelt:1987im,Raffelt:1996wa,kuster2007axions}:
	\begin{equation}
	\left[\omega_\gamma+i \partial_z + \left(
	\begin{array}{ccc}
		\Delta _\perp & \Delta _R & 0 \\
		\Delta _R & \Delta _{\parallel} & \Delta_{a\gamma} \\
		0 & \Delta_{a\gamma}  & \Delta _a \\
	\end{array}
	\right) \right] \left(
	\begin{array}{c}
		\gamma_\perp \\
		\gamma_\parallel \\
		a \\
	\end{array}
	\right)=0\;,
	\label{eq:eom3}
\end{equation}
where, $\omega_\gamma$ is the energy of the photon. 
The two states $\gamma_{\perp}$ and $\gamma_{\parallel}$ denote the photon polarization states perpendicular and parallel to the transverse magnetic field $\vec{B}_T$ (i.e. the component of magnetic field perpendicular to the direction of propagation), while $a$ denotes the ALP. 
The diagonal terms $\Delta_\perp$ and $\Delta_\parallel$ come from the effective masses of the two photon polarization states in the plasma, while $\Delta_a$ is the contribution from ALP mass.
The off-diagonal term $\Delta_R$ represents the contribution from Faraday rotation, which is regulated by the magnetic field parallel to the direction of propagation.
The ALP-photon mixing is generated by the $\Delta_{a\gamma}$ term, which is regulated by the transverse magnetic field $\vec{B}_T$.

The Faraday rotation term would be significant near the magnetar, leading to all three states ($\gamma_{\perp}$, $\gamma_{\parallel}$, and $a$) mixing with each other. However, away from the magnetar, in the interstellar medium, the Faraday rotation effects can be neglected. Therefore, in this region, we can simplify the problem to the two-state basis $(\gamma_\parallel, \,a)$. The ALP-photon effective Hamiltonian in this basis is
\begin{equation}
	\mathcal{H}_{\text{eff}} = -
	\begin{pmatrix} 
		\Delta_{\text{pl}} & \Delta_{a\gamma} \\
		\Delta_{a\gamma} & \Delta_a \\
	\end{pmatrix}
\; .
\label{eq:Heff}
\end{equation}
Where the individual elements of the matrix are~\cite{Raffelt:1987im,Raffelt:1996wa,kuster2007axions}
\begin{equation}
\Delta_a=-\frac{m_a^2}{2\, \omega_\gamma} \; , \qquad \Delta_{\text{pl}} =\Delta_{\parallel}=-\frac{m_{\text{eff}}^2}{2\, \omega_\gamma}\;, \qquad \Delta_{a\gamma}= \frac{g_{a\gamma}|\vec{B}_T| }{2}  \;.
\end{equation}
Here, $m_a$ is the mass of the ALP, while $g_{a\gamma}$ is the coupling between photons and ALPs.
The effective mass of the photon, denoted by $m_{\text{eff}}$, can be expressed as~\cite{Dobrynina:2014qba}
\begin{equation}
	m^2_{\text{eff}} = \omega_{\text{\text{pl}}}^2 - m_{\text{EM}}^2\;,
	\label{eq:meffsq00}
\end{equation}
where the plasma frequency $\omega_{\text{pl}}$ ~\cite{Braaten:1993jw} and the electromagnetic (EM) contribution $m_{\text{EM}}^2$ are\footnote{For a detailed discussion on how the EM contribution is modified due to the nature of the magnetic field near a magnetar, see~\cite{Millar:2021gzs,Bondarenko:2022ngb}.  For our purpose, the approximation made above is sufficient.}
\begin{equation}
	 \omega_{\text{pl}}^2 = \frac{4\pi \alpha \, n_e}{ m_e }\;, \qquad m_{\text{EM}}^2 = \frac{88 \, \alpha^2 \omega_\gamma^2}{135 \,m_e^4} \rho_{\text{EM}}\;.
	 \label{eq:meffsq01}
\end{equation}
Here, $\alpha$ is the fine structure constant, $n_e$ is the electron density in the medium, and $m_e$ is the mass of the electron. Note that the plasma frequency predominantly depends upon the electrons --- the lightest charged particles --- in the medium. This is because the plasma frequency is inversely proportional to the mass of the charged particle. The EM energy density $\rho_{\text{EM}}$ in the presence of strong magnetic fields can be written as $\rho_{\text{EM}} = B^2/2$. Except near regions like magnetar neighborhoods, where the magnetic field is extremely strong, the contribution from the $m_{\text{EM}}^2$ term will be negligible.

Typically in the interstellar medium, $\Delta_{a\gamma} \ll | \Delta_{\text{pl}} - \Delta_a |$, which leads to a small mixing angle and therefore small value for the photon-ALP conversion.
On the other hand, in the magnetar neighborhood, it is possible that the resonance condition $\Delta_{\text{pl}}=\Delta_a$ would be fulfilled, in which case the mixing angle would be very large. The conversion probabilities in this resonance scenario, corresponding to $m^2_{\text{eff}} =m_a^2$, need to be treated separately.

\subsection{Without Resonance}

The $\gamma_\parallel$ component of an emitted photon couples to ALPs and hence will undergo oscillations in the presence of magnetic field.
In the two-state approximation, the evolution of state $(\gamma_\parallel,\, a)$ is represented by the differential equation
\begin{equation}
	i \frac{d}{dl} 	\left(\begin{array}{c}
		\gamma_\parallel \\
		a \\
	\end{array}\right)
 = \mathcal{H}_{\text{eff}} \left(\begin{array}{c}
 	\gamma_\parallel \\
 	a \\
 \end{array}\right)\;.
\end{equation}
The state after traversing a distance $l$ is then given by
\begin{equation}
	\left(\begin{array}{c}
		\gamma_\parallel (l) \\
		a (l)\\
	\end{array}\right)
=\exp \left[ -i \int_{0}^{l} \mathcal{H}_{\text{eff}} \, dl^\prime \right] 	\left(\begin{array}{c}
	\gamma_\parallel (0) \\
	a (0)\\
\end{array}\right) \; ,
\end{equation}
where $\mathcal{H}_{\text{eff}}$ is the effective Hamiltonian from Eq.~(\ref{eq:Heff}).
The conversion probability $P (\gamma \rightarrow a)$ is given by $P (\gamma \rightarrow a) \equiv \left|\langle a(l) | \gamma_\parallel (0)\rangle \right|^2$.
In the limit of constant electron density and magnetic field, and hence constant $\mathcal{H}_{\text{eff}}$,  the photon-ALP conversion probability is
\begin{equation}
	P (\gamma \rightarrow a)
	= 4 \left(\dfrac{\Delta_{a\gamma}^2 }{\Delta_{\text{osc}}^2}\right) \sin^2 \left(  \frac{\Delta_{\text{osc}} \, l}{2} \right)\;,
\end{equation}
with $P (a \rightarrow \gamma)= P (\gamma \rightarrow a)$. Here the oscillation wavenumber $\Delta_{\text{osc}}$ is given by
\begin{equation}
	\Delta_{\text{osc}}^2=(\Delta_{pl}-\Delta_a)^2+4\, \Delta_{a\gamma}^2\;.
\end{equation}
For propagation in the interstellar medium, we choose the limit of constant electron density and magnetic field, which allows us to easily estimate the behavior of the conversion probability.
Note that this probability, regulated by $\Delta_{a\gamma}^2 / \Delta_{\text{osc}}^2$, would be small in the limit $\Delta_{a\gamma} \ll | \Delta_{\text{pl}} - \Delta_a |$.

\subsection{With Resonance}
As a photon propagates outward from the surface of a magnetar, the effective mass of the photon changes due to its dependence on the magnetic field and electron density. This may lead to fulfillment of the resonance condition $m^2_{\text{eff}} =m_a^2$. To investigate the effects of resonance on the transition probability, we first calculate the Landau-Zener adiabaticity parameter $\ad_{\text{ad}}$~\cite{landau1932theory,Majorana:1932ga,stuckelberg1932theory,Zener:1932ws}, defined at the point of resonance $l_i$ as \footnote{See~\cite{Carenza:2023nck} for caveats and detailed discussions concerning the applicability of the Landau-Zener formula.}
\begin{equation}
	\ad_{i} \equiv  \, \left[ \dfrac{ 4\, \Delta_{a\gamma}^2}{ \left| d \Delta_{\text{pl}}/ dl  \right| } \right]_{l=l_i}
	=  \frac{2 \, g_{a\gamma}^2 \, \omega_\gamma}{m_a^2} \, B_{T,i}^2\, \mathcal{R}_i\; ,
\end{equation}
where we have defined
\begin{equation}
	B_{T,i} \equiv B_T (l_i)\;, \quad \mathcal{R}_i \equiv \Big| \frac{d \ln m^2_{\text{eff}}}{
		dl} \Big|_{l=l_i}^{-1}\; .
\end{equation}
For a single resonance, the flip probability $\mathbb{P}_{\text{flip}\, , i}$ between the mass eigenstates can be approximated as
\begin{equation}
	\mathbb{P}_{\text{flip},\, i} \approx e^{-\pi \ad_{i} / 2}\;.
\end{equation}
The flip probability $\mathbb{P}_{\text{flip},\, i} $ may be interpreted as the probability $\mathbb{P}_i (\gamma \to \gamma) $ survival probability if far away from resonance, the off-diagonal mixing term in $\mathcal{H}_{\text{eff}}$ is much smaller than the difference between the diagonal terms. That is, for $\left| l-l_i \right| \gg 0$, we need to have
\begin{equation}
	 2 \, |g_{a\gamma} \, B_{T}  (l) \, \omega_\gamma| \ll \left| m_{\text{eff}}^2 (l) - m_a^2 \right|\;.
	 \label{eq:rescond}
\end{equation}
This condition is fulfilled near the surface of the magnetar, due to the large values of $|m_{\text{eff}}^2|$, and also very far away from the magnetar, due to the very small values of $B_{T} (l)$. If this condition is satisfied around a resonance, it would imply that the survival probability $\mathbb{P} (\gamma \to \gamma)$ at that resonance `$i$' would be
\begin{equation}
	\mathbb{P}_i (\gamma \to \gamma) = e^{-\pi \ad_{i} / 2}\;.
\end{equation}
The conversion probability for a single resonance is given by
\begin{equation}
		\mathbb{P}_i (\gamma \to a) = 1 - e^{-\pi \ad_{i} / 2}\; .
\end{equation}
For small $\ad_{i}$ values, this can be approximated as
\begin{equation}
	\mathbb{P}_i (\gamma \to a) \approx \pi \ad_{i} / 2\; .
	\label{eq:singleressmall}
\end{equation}
\subsection{With a Pair of Resonances}
Due to large magnetic fields typically present close to the magnetars,  $m_{\text{EM}}^2$ contribution  in Eq.~(\ref{eq:meffsq00}) dominates over the $\omega_{\text{pl}}^2$ contribution for X-ray photons. This results in $m^2_{\text{eff}} <0$, and consequently  $m^2_{\text{eff}} <m_a^2$ in this region. Also, far away from the magnetar $m^2_{\text{eff}} \to 0$ due to the decreasing magnetic field and electron density, resulting again in $m^2_{\text{eff}} <m_a^2$. The flavor conversions will be significant if there are resonances in the intermediate region, where $m^2_{\text{eff}} $ crosses $m_a^2$. The above arguments show that the number of such resonances must always be even. Even when there are multiple resonance along the way, the resonances can be split up into pairs where $m^2_{\text{eff}}$ crosses $m_a^2$ from below at the first resonance and from above in the resonance immediately after. 

We therefore first focus on one such pair. The photon survival and conversion probabilities while crossing such a pair can be expressed as
\begin{align}
	\mathbb{P} (\gamma \to \gamma) =&\, P_1 P_2 + \left(1- P_1  \right) \left(1- P_2 \right)\;, \\
	\mathbb{P} (\gamma \to a) = & \, P_1 \left( 1-P_2 \right) + P_2 \left( 1-P_1 \right)\;,
	\label{eq:tworesexact}
\end{align}
where $P_i \equiv \mathbb{P}_i (\gamma \to a) = 1- e^{-\pi \ad_{i} / 2}$ is the conversion probability at the resonance `$i$'.
Note that if $P_1$ and $P_2$ are small, the net conversion probability for this pair would become
\begin{align}
	\mathbb{P} (\gamma \to a) = & \, P_1 \left( 1-P_2 \right) + P_2 \left( 1-P_1 \right) = \, P_1 +P_2 - 2 P_1 P_2
	\approx \,P_1 +P_2 \; .
	\label{eq:twores}
\end{align}
As a photon starts propagating from the surface of the magnetar, the resonance condition ($m^2_{\text{eff}} =m_a^2$) is likely to be fulfilled naturally multiple times in the magnetar neighborhood. This happens because, while Eq.~(\ref{eq:Bspatial}) and Eq.~(\ref{eq:nspatial}) capture the overall behavior, the background electron density and the magnetic field strength typically undergo fluctuations.
In the next section, we explore the effects of multiple such pairs of resonances in succession.

\subsection{With Multiple Resonances and Different Domains}
\label{sec:multres}

To calculate the photon to ALP conversion probability in the presence of many fluctuations, we need to to know the overall behavior of the environment.
We assume that, as the photon propagates outwards from the surface of the magnetar, in addition to fluctuations in the electron density and magnetic field strength, it also encounters many domains with magnetic fields oriented randomly towards an arbitrary direction. Further, due to the randomness of orientation of the magnetic field, we assume that the angles at which each domain are misaligned with the next one are $\sim O(1)$.

We denote the intensities of the two photon polarizations (defined, for the sake of definiteness, as $\gamma_1 \equiv \gamma_\perp (0)$, $\gamma_2 \equiv \gamma_{\parallel}(0)$ at the first domain) and the ALP by $I_{\gamma_1,\, \gamma_2}$, and $I_a$, respectively.
The random orientation of the magnetic field in each domain ensures that both the photon polarizations mix equally with the ALPs on an average.
If $p_n$ denotes the probability of conversion in the $n$-th domain, then we get the recursion relation
\begin{equation}
	\left(\begin{array}{c}
		I_{\gamma_1} (n) \\[5pt]
		I_{\gamma_2} (n) \\[5pt]
		I_a (n) \\
	\end{array}\right) =  
	\left(\begin{array}{ccc}
		1- \frac{1}{2}p_{n} & 	0& 	\frac{1}{2} p_{n} \\[5pt]
	0 & 	1- \frac{1}{2}p_{n} & 	\frac{1}{2} p_{n} \\[5pt]
		\frac{1}{2} p_{n}&	\frac{1}{2} p_{n}&1- p_{n}
	\end{array}\right)
	\left(\begin{array}{c}
	I_{\gamma_1} (n-1) \\[5pt]
	I_{\gamma_2} (n-1) \\[5pt]
	I_a (n-1) \\
	\end{array}\right)  \;.
\end{equation}
 Here we ignore possible interferences between resonances, i.e., phase effects~\cite{Dasgupta:2005wn}.
Defining $I_\gamma \sim I_{\gamma_1}+ I_{\gamma_2}$, one may write the above equation as~\cite{Grossman:2002by}
\begin{align}
		\left(\begin{array}{c}
		I_\gamma (n) \\[5pt]
		I_a (n) \\
	\end{array}\right) =  &
\left(\begin{array}{cc}
	1- \frac{1}{2}p_{n} & p_{n} \\[10pt]
	\frac{1}{2} p_{n} & 1 - p_{n} \\
\end{array}\right)
	\left(\begin{array}{c}
	I_\gamma (n-1) \\[5pt]
	I_a (n-1) \\
\end{array}\right) \;.
\end{align}
The solution to the above recursion relation may be written as (see Appendix~\ref{sec:appendix})
\begin{align}
		\left(\begin{array}{c}
	I_\gamma (n) \\[5pt]
	I_a (n) \\
\end{array}\right)	= &\left(\begin{array}{cc}
		\frac{2}{3} \left[ 1+\frac{1}{2} \prod_{k=1}^{n} \left(1 - \frac{3}{2} p_k \right) \right] &  \quad 	\frac{2}{3} \left[ 1- \prod_{k=1}^{n} \left(1 - \frac{3}{2} p_k \right) \right]  \\[10pt]
		\frac{1}{3} \left[ 1 - \prod_{k=1}^{n} \left(1 - \frac{3}{2} p_k \right) \right] & \quad
		\frac{2}{3} \left[ \frac{1}{2} + \prod_{k=1}^{n} \left(1 - \frac{3}{2} p_k \right) \right]  \\
	\end{array}\right) 
	\left(\begin{array}{c}
		I_\gamma (0) \\[5pt]
		I_a (0) \\
	\end{array}\right)\; .
\end{align}
Therefore, the photon to ALP conversion probability in the presence of multiple resonances can be expressed as
\begin{equation}
	\mathbb{P} (\gamma \to a) = \frac{1}{3} \left[ 1-\prod_{k=1}^{n_d} \left(1 - \frac{3}{2} \, p_k \right)  \right] \;,
\end{equation}
where $n_d$ is the number of domains. The above equation, in the special case of all $p_k$'s having the same value, matches with~\cite{Grossman:2002by}. In the limit of large number $n_d$ of domains, this can be approximated as the limiting value
\begin{equation}
		\mathbb{P} (\gamma \to a) = \frac{1}{3} \left[ 1-\exp \left( - \frac{3}{2} \sum_{k=1}^{n_d}p_k \right)  \right] \; .
		\label{eq:ptot0}
\end{equation}
Note that, each of the domain may have zero or a finite number of resonances. Therefore, in the approximation that the dominant contribution to flavor conversions are near resonances, we obtain
\begin{equation}
	\sum_{k=1}^{n_{d}}p_k =\sum_{i=1}^{n_{r}} P_i \;.
	\label{eq:sumPi}
\end{equation}
Here, the index `$k$' counts the number of domains, whereas the index `$i$' counts the number of resonances. The total number of resonances is denoted by $n_r$.
The smallness of the individual $P_i$'s allows us to approximate
\begin{equation}
	\sum_{i=1}^{n_{r}} P_i \approx \sum_{i=1}^{n_{r}} \frac{\pi \ad_i }{2} = \frac{\pi \ad_{\text{tot}}}{2}\; ,
	\label{eq:gammatot}
\end{equation}
where $\ad_{\text{tot}}$ is given by
\begin{equation}
\ad_{\text{tot}}	= \frac{2 \, g_{a\gamma}^2 \, \omega_\gamma}{m_a^2} \sum_{i}^{n_{r}}  B_{T,i}^2\mathcal{R}_i\; .
\end{equation}
Therefore, we can estimate the total conversion probability to be
\begin{equation}
	P_{\text{tot}} \approx \frac{1}{3} \left( 1 - e^{-\frac{3 \pi}{4} \ad_{\text{tot}}} \right)\; .
	\label{eq:ptot}
\end{equation}
Interestingly, in the small $\ad_{\text{tot}}$ limit, we obtain $P_{\text{tot}} \approx \frac{1}{2}  \frac{\pi \ad_{\text{tot}}}{2} = \frac{1}{2} \sum P_i$. This is because, for completely unpolarized light, only half of the photons couple to ALPs at any given point and hence experience resonance. On the other hand, due to the existence of either many domains or a few domains with large $\ad_{i}$ values, in the scenario where the value of $\ad_{\text{tot}}$ becomes large, Eq.~(\ref{eq:ptot}) gives $P_{\text{tot}} \sim 1/3$. This is expected, as the intensities of the photon polarizations and the ALP would be driven towards full depolarization of all the three states in this scenario.

In section~\ref{sec:probabilities} we calculate the $\gamma \to a$ conversion probability in the magnetar neighborhood and $a \to \gamma$ conversion probability in the ISM.

\section{\label{sec:probabilities}Photon-ALP-Photon Conversions During Propagation}
For the purpose of this work, we need to estimate the number of photons, emitted from the surface of the magnetar, that may cross the opaque barrier by oscillating into ALPs.
To calculate this  $\gamma \to a \to \gamma$ oscillation probability, we need to calculate the conversion probability $P(\gamma \to a)$ in the magnetar neighborhood and $P(a\to \gamma)$ in the ISM.
In this section, we discuss in detail the key features of the conversion probabilities and quantify their dependence on different input parameter values.

\subsection{Locations of the Resonances in the Magnetar Neighborhood}
For a photon emitted from a magnetar, the resonance condition ($\Delta_{\text{pl}}=\Delta_a$) is likely to be met several times due to the turbulent nature of the potential caused by the varying electron density and magnetic field. However, to start with, we shall make the assumption that these fluctuations are absent and investigate the effect of resonances in the magnetar neighborhood.

We first estimate the distance at which the resonance happens.
The resonance condition is satisfied when $m^2_{\text{eff}} =m_a^2$.
The spatial behavior of the $m^2_{\text{eff}}$ may be obtained from Eqs.~(\ref{eq:Bspatial}),~(\ref{eq:nspatial}),~(\ref{eq:meffsq00}), and (\ref{eq:meffsq01}) as
\begin{equation}
	m^2_{\text{eff}} \equiv \frac{4 \pi  \alpha }{m_e} \, n_0 \left( \frac{r}{r_0} \right)^{-3}  - \frac{88\, \alpha^2}{270 \, m_e^4} \, B_0^2\, \omega_\gamma^2 \left( \frac{r}{r_0} \right)^{-6}\;,
	\label{eq:meffsqspatial0}
\end{equation}
where we have assumed that the power law dependence $n_e \sim r^{-3}$, as in Eq.~(\ref{eq:nspatial}). We redefine the coefficients as
\begin{equation}
	c_n \equiv \frac{4 \pi  \alpha }{m_e}  \; , \qquad c_B \equiv \frac{88\, \alpha^2}{270 \, m_e^4} \;.
\end{equation}
This allows us to write the following condition at the point of resonance:
\begin{equation}
	m^2_{\text{eff}}  = c_n \, n_0 \, \left( \frac{r}{r_0} \right)^{-3}  - c_B \, B_0^2\, \omega_\gamma^2 \, \left( \frac{r}{r_0} \right)^{-6} \, = \, m_a^2\;.
	\label{eq:meffsqspatial}
\end{equation}
This condition is satisfied for $r=r_{\lessgtr}$ such that
\begin{equation}
	 r_{\lessgtr} = r_0 \left[ \frac{2 \, c_B \,  \omega_\gamma^2 \, B_0^2}{c_n \, n_0 \pm \sqrt{ c_n^2 \, n_0^2 - 4 \, c_B \, B_0^2 \; \omega_\gamma^2 \, m_a^2 } }   \right]^{1/3} \; .
	 \label{eq:nearfarres}
\end{equation}
Note that the `$+$' sign leads to smaller values, which corresponds to the solution $r_{<}$, i.e., the near resonance point, whereas the `$-$' sign (leading to larger values) corresponds to the solution $r_{>}$, i.e., the far resonance point.
For $c_n^2\, n_0^2 < 4 \, c_B B_0^2\, \omega_\gamma^2\, m_a^2 \,$, there is no real solution of $r_{\lessgtr}$. This is because in this case the mass of the ALP is always more than the effective mass of the photon, i.e.,
\begin{equation}
	m_a^2 \, > \, \left[ m_{\text{eff}}^2 \right]_{\max}\;,
\end{equation}
 and hence the resonance condition  can never be fulfilled.
 
 \begin{figure}[t]
 	\centering
 	\includegraphics[width=.55 \textwidth]{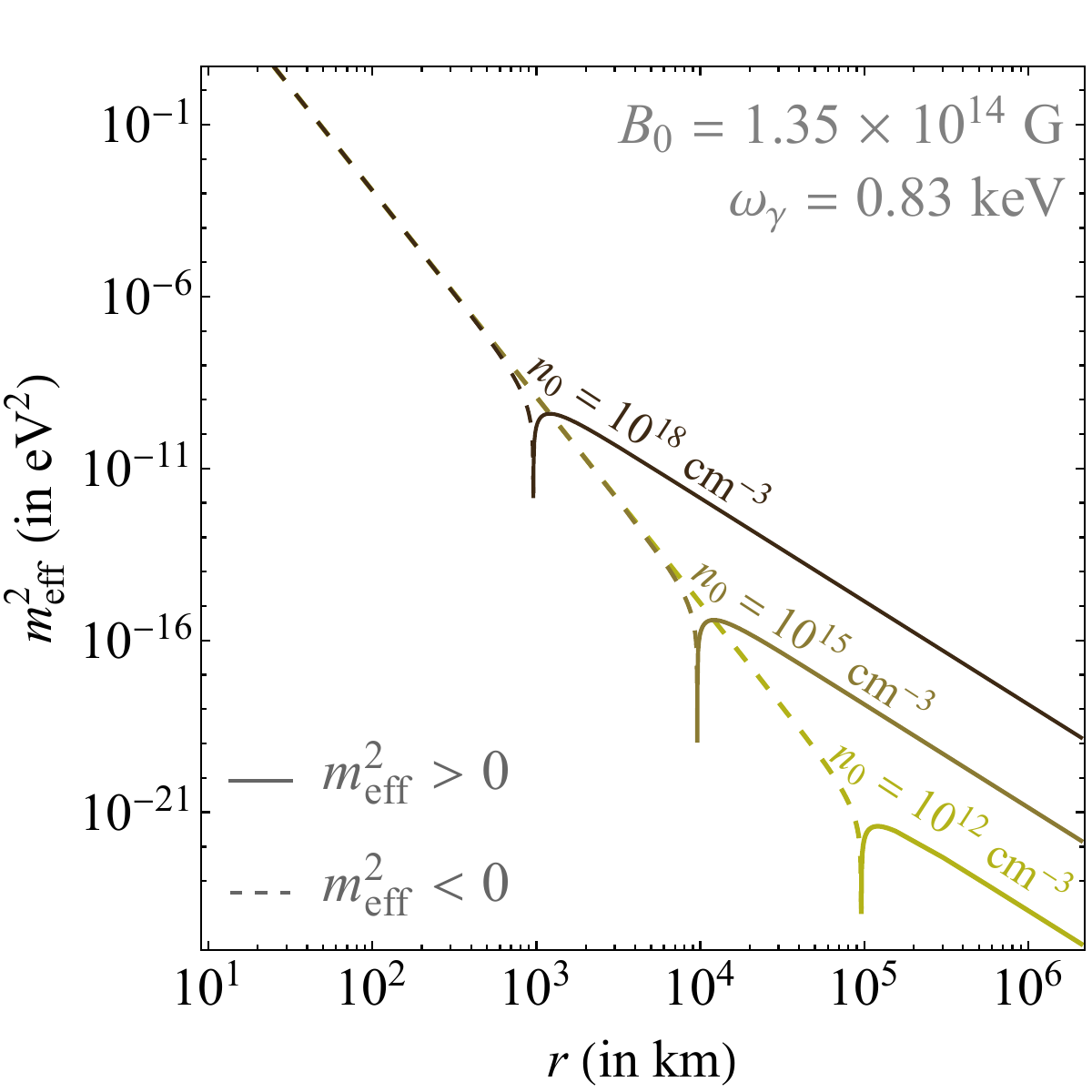}
 	\caption{
 		Behavior of $m^2_{\text{eff}}$ (in eV$^2$) vs. $r$ (in km), for different values of the surface electron density $n_0$. Solid (dashed) lines represent $m^2_{\text{eff}} > 0$ ($m^2_{\text{eff}} < 0$) values. The resonance condition $m^2_{\text{eff}} = m_a^2$ cannot be satisfied at $m^2_{\text{eff}} < 0$ values. When $m^2_{\text{eff}} > 0$ and $m_a^2 < \left[m_{\text{eff}}^2 \right]_{\max}$, the resonance condition would be satisfied twice.The $B_0$ and $\omega_\gamma$ values shown correspond to the magnetic field strength and energy that will be relevant for the magnetar PSR J1622-4950.}
 	\label{fig:resonance}
 \end{figure}
 In Fig.\,\ref{fig:resonance}, we plot the spatial behavior of $m_{\text{eff}}^2$ as a function of the distance from the center of the magnetar for different $n_0$ values. 
The photons emitted from the surface of the magnetar of radius $\sim 10$ km initially have a negative $m^2_{\text{eff}}$ value due to the contribution from the $B_0^2$ term in Eq.~(\ref{eq:meffsqspatial0}). As the photons move outward from the magnetar, this contribution decreases faster than the positive contribution due to the $n_0$ term in Eq.~(\ref{eq:meffsqspatial0}), and hence the negative value of $m^2_{\text{eff}}$ moves closer to zero. At sufficiently large $r$, the value of $m^2_{\text{eff}}$ crosses zero, further attains a maximum and later decreases towards zero. From Fig.\,\ref{fig:resonance}, we observe the following:
\begin{itemize}
	\item If the maximum value of $m^2_{\text{eff}}$ is greater than $m_a^2$, the resonance condition will be satisfied twice. 
	For example, if $m_a^2 = 10^{-18}$ eV$^2$ (i.e., $m_a= 10^{-9}$ eV) and $n_0 = 10^{15} \text{ cm}^{-3}$, resonances would occur at $r_< \approx 10^4$~km and $r_> \approx 10^5$ km.
	\item  On the other hand, if $m^2_{\text{eff}}$ cannot reach $m_a^2$, then there will be no resonances. For example, in the range $n_0 \sim (10^{12} - 10^{18}) \text{ cm}^{-3}$, the resonance condition can never be fulfilled for $m_a^2 \gtrsim 10^{-10}$ eV$^2$ (i.e., $m_a \gtrsim 10^{-5}$ eV).
	\item  With increasing $n_0$, the maximum value of $m^2_{\text{eff}}$ increases and hence the resonance condition can be satisfied for a wider range of $m_a$. Moreover, increasing $n_0$ leads to  $m^2_{\text{eff}}$ values becoming positive at a  lower value of $r$, thus bringing the near resonance point $r_<$ closer.
\end{itemize}

\begin{figure}[!t]
	\centering
	\includegraphics[width=.46 \textwidth]{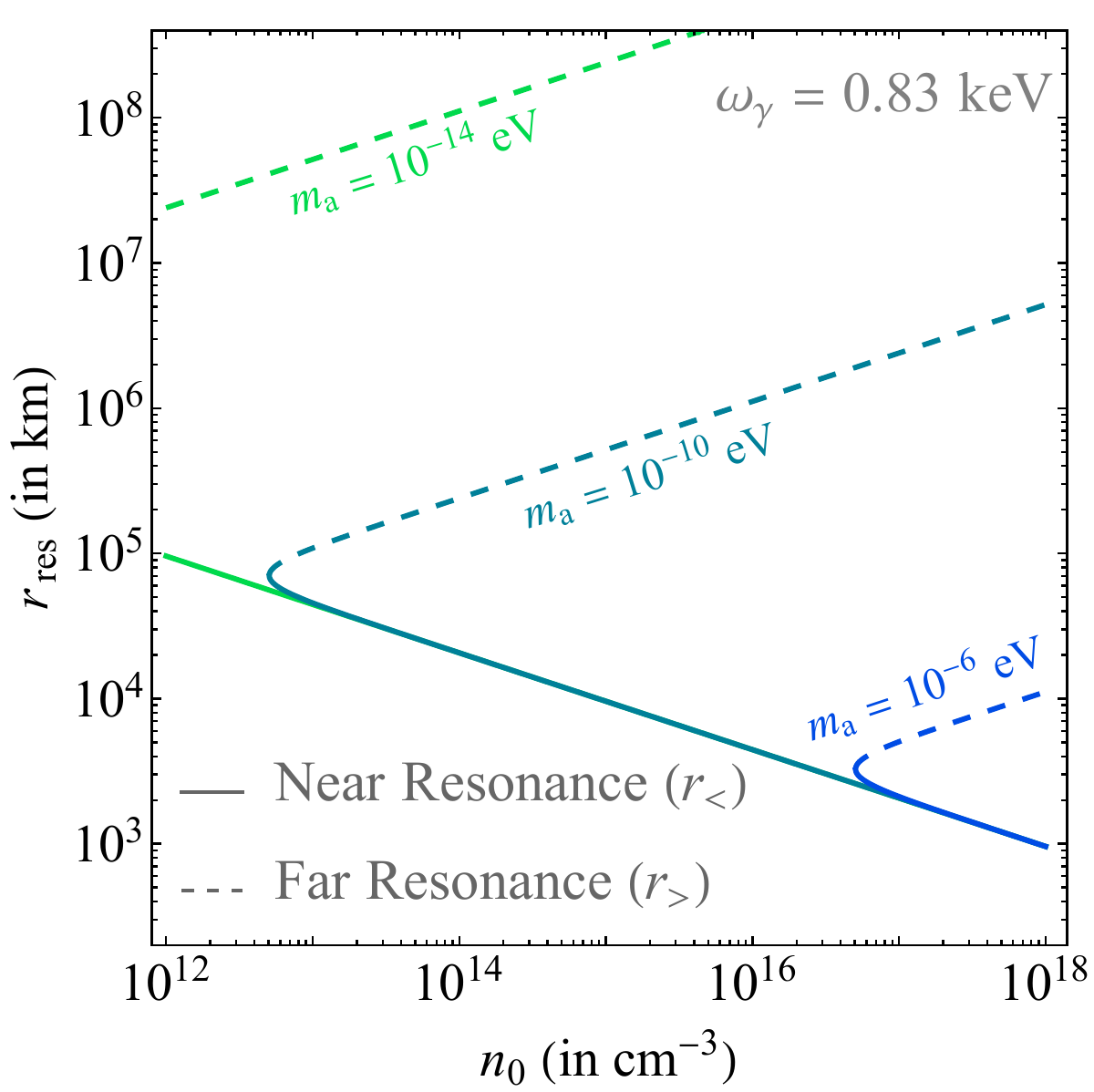}
	\hspace{2pt}
	\includegraphics[width=.48 \textwidth]{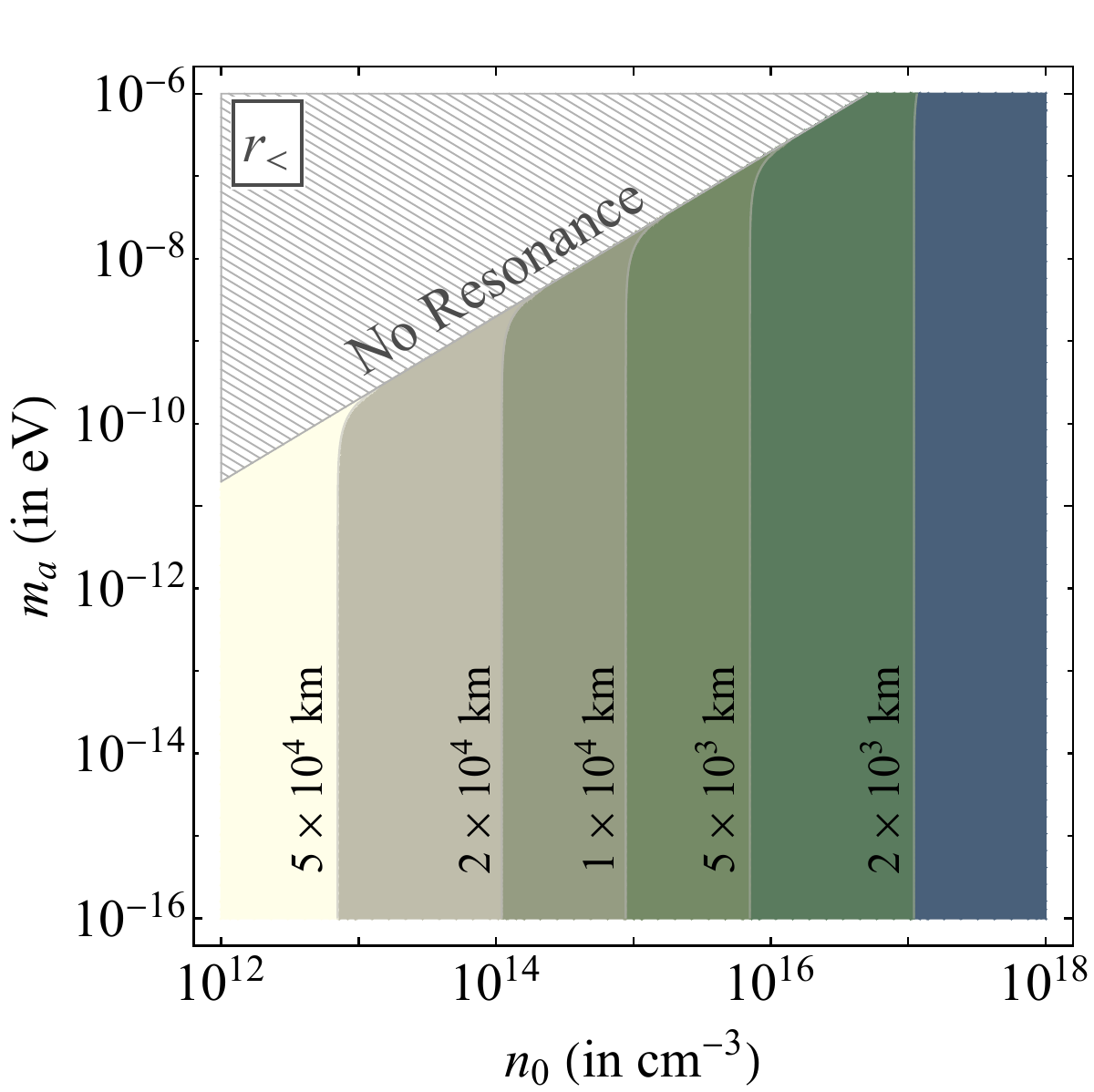}
	\caption{Left panel: the  positions of the near and far resonance points $r_<$ and $r_>$ as functions of $n_0$, for different values of $m_a$. The near (far) resonance points lie along the solid (dashed) lines. The turnaround point for each $m_a$ curve corresponds to the minimum value of $n_0$ for which resonance can happen.
	Right panel: the position of the near resonance point $r_<$ in the $(n_0,\, m_a)$ plane. The gray hatched region represents parameter values for which resonance condition is not satisfied.}
	\label{fig:resonancecontour}
\end{figure}

In Fig.~\ref{fig:resonancecontour} (left panel), we show the positions of the near and far resonance points  $r_<$ and $r_>$ as functions of $n_0$, for different values of $m_a$.
 We also observe the increase in $n_0$ leading to a smaller value of $r_<$. The position of the near-resonance point does not depend significantly on the value of $m_a^2$ due to the sharp nature of the zero-crossing, as observed in Fig.\,\ref{fig:resonance}, and immediately reaching the maximum $m^2_{\text{eff}}$ value. Although too large $m_a^2$ values do not meet the resonance criteria, leading to the V-like turnaround behavior.

The far resonance point depends inversely (compared to the near resonance point) to the value of $n_0$, as a higher value of $n_0$ leads to a higher contribution to $m^2_{\text{eff}}$ at further distances, which leads to the far resonance condition being satisfied further away. The far resonance point also depends upon the mass of the ALP, as a higher mass meets the $m^2_{\text{eff}}=m_a^2$ condition sooner, whereas for lower $m_a$ values, the far resonance point $r_{>}$ shifts further away from the magnetar.

The dependence of the near and far resonance points on the parameters $m_a$ and $n_0$ can be approximated using Eq.~(\ref{eq:nearfarres}). In the presence of resonance, the condition $c_n^2 \, n_0^2 > 4 \, c_B B_0^2 \, \omega_\gamma^2 \, m_a^2\,$ must be satisfied. Therefore, for small values of $m_a$, such that the condition $c_n^2 \, n_0^2 \gg 4 \, c_B B_0^2 \, \omega_\gamma^2 \, m_a^2 \,$ is satisfied, we can expand and rewrite Eq.~(\ref{eq:nearfarres}) as
\begin{equation}
		 r_{\lessgtr} \approx r_0 \left[ \frac{2 \, c_B \,  \omega_\gamma^2 \, B_0^2}{c_n \, n_0 \pm c_n \, n_0 \left(1 - \frac{  2 \, c_B B_0^2 \, \omega_\gamma^2 \, m_a^2}{c_n^2 \, n_0^2} \right) }   \right]^{1/3} \;.
\end{equation}
For near resonance $r_<$, which corresponds to the `$+$' sign in the denominator, this gives
\begin{equation}
	r_< \approx r_0 \left[ \frac{\, c_B \,  \omega_\gamma^2 \, B_0^2\,}{c_n \, n_0}   \right]^{1/3} \;.
	\label{eq:rlessapprox}
\end{equation}
Thus pointing out the lack of dependence of $r_<$ on the value of $m_a$, and the power law dependence of $r_< \propto n_0^{-1/3}$. On the other hand, for the far resonance $r_>$, which corresponds to the `$-$' sign in the denominator, we get
\begin{equation}
	r_> \approx r_0 \left[ \frac{c_n \, n_0}{m_a^2 }   \right]^{1/3} \; ,
	\label{eq:rgtrapprox}
\end{equation}
thus, revealing the dependence of $r_>$ on both $m_a$ and $n_0$.

In the contour plot in Fig.\,\ref{fig:resonancecontour} (right panel), we show the dependence of the position of $r_<$ as a function of $n_0$ and $m_a$. This further emphasizes the point that the location of the near-resonance is independent of the mass of the ALP except near the turnaround point. However, note that the existence of the near-resonance is dependent upon the mass, as not all $m_a$ values may be accessible, depending on the maximum effective mass of the photons in the magnetar neighborhood. The parameter ranges for which the resonance criteria is not satisfied is denoted by the gray hatched region.

\subsection{The Conversion Probability $P_{MN}$ in the Magnetar Neighborhood}
\label{sec:pMagnetar}

First, we take the idealized scenario where we do not consider fluctuations in magnetic field and electron density. We assume that the behavior of the transverse magnetic field and the electron density is as described in Eq.~(\ref{eq:Bspatial}) and (\ref{eq:nspatial}).
While passing through a resonance, since $m^2_{\text{eff}} \approx m_a^2$, the adiabaticity parameter $\ad_{i} \equiv \ad (r_i)$ corresponding to a single resonance can be simplified to the form
\begin{equation}
\ad (r_i) = 2 \, g_{a\gamma}^2 \, \omega_\gamma \, B_{T,i}^2 \, \Big| \frac{d m^2_{\text{eff}}}{dr} \Big|_{r_i}^{-1} \; ,
\end{equation}
where the resonance point is denoted as $r_i$. The inverse relationship of $\ad (r_i) $ with the derivative term will be instrumental in estimating the general power law behavior of the conversion probability at the near and far resonance points.\\

\noindent
\textbf{\textit{At the near-resonance point}}: The near resonance conditions are satisfied around the region where $m_{\text{eff}}^2$ becomes greater than zero. This means that both the competing contributions --- regulated by $n_0$ and $B_0$ --- are important here. This warrants a careful scrutiny of the power-law behavior of each of the contributions.
The derivative of $m_{\text{eff}}^2$ is
\begin{align}
	\frac{d m^2_{\text{eff}}}{dr} & =  - \frac{3}{r}   \, c_n  \, n_0 \left( \frac{r}{r_0} \right)^{-3}   + \frac{6}{r} \, c_B B_0^2\, \omega_\gamma^2 \left( \frac{r}{r_0} \right)^{-6} \nonumber \\
	& =  - \frac{3}{r} m^2_{\text{eff}} + \frac{3}{r} \, c_B B_0^2\, \omega_\gamma^2 \left( \frac{r}{r_0} \right)^{-6} \;.
\end{align}
At the near-resonance point $r_<$, this gives
\begin{equation}
	\left. \frac{d m^2_{\text{eff}}}{dr}  \right|_{r_<}  = - \frac{3}{r_<}  m^2_a + \frac{3}{r_0} \, c_B B_0^2\, \omega_\gamma^2 \left( \frac{r_<}{r_0} \right)^{-7}\; .
\end{equation}
Therefore, in the limit $m_a \to 0$, the primary power-law behavior of $d m^2_{\text{eff}}/ dr$ can be simply expressed as
\begin{equation}
	\frac{d m^2_{\text{eff}}}{dr} \Big|_{r_<}  \propto \;r_<^{-7}\;.
\end{equation}
Since $B_T \propto r^{-3}$, we get the power-law dependence of $\ad (r_< )$ to be
\begin{equation}
 \ad (r_< )\; \propto \; B_{T,<}^2 \left| \frac{d m^2_{\text{eff}}}{dr} \right|_{r_<}^{-1} \; \propto \; r_<\;. 
\end{equation}
As discussed previously (and observed in Fig.\,\ref{fig:resonance} and Fig.\,\ref{fig:resonancecontour} [left]), an increase in $n_0$ would shift the near-resonance point $r_<$  closer to the magnetar surface. 
This decrease in $r_<$ would also decrease $\ad_{\text{tot}}$ and hence  $P_{\text{tot}}$.
On the other hand, a decrease in $n_0$ would result in the increase of $P_{\text{tot}}$.
As discussed in the previous section, dependence of $r_<$ on $m_a$ is negligible, leading to almost no dependence of $m_a$ on $P_{\text{tot}}$.\\

\noindent
\textbf{\textit{At the far-resonance point}}: At the far-resonance point $r_>$, the dominant contribution to $m^2_{\text{eff}}$ comes from the $n_0$ term. This is because the $B_0^2$ term falls as $r^{-6}$ compared to the $n_0$ term which falls as $\sim r^{-3}$. Therefore, the derivative term can be approximated as
\begin{equation}
	\frac{d m^2_{\text{eff}}}{dr}  \approx - \frac{3}{r_0}   \, c_n  \, n_0 \left( \frac{r}{r_0} \right)^{-4} \;.
\end{equation}
This implies that, at this point the power-law dependence of $\ad\, (r_>)$ is
\begin{equation}
	\ad\, (r_>) \; \propto \; B_{T,>}^2 \left| \frac{d m^2_{\text{eff}}}{dr} \right|_{r_>}^{-1} \; \propto \; r_>^{-2}\;.
	\label{eq:farresPL}
\end{equation}
As discussed previously, an increase in $n_0$ would mean that the far-resonance $r_>$ occurs further away from the magnetar, as it will take longer for $m^2_{\text{eff}}$ to reach the value of $m_a^2$. From Eq.~(\ref{eq:farresPL}), this would lead to a decrease in $\ad_{\text{tot}}$, and thereby a decrease in $P_{\text{tot}}$. On the other hand, a decrease in $n_0$ would lead to an increase in $P_{\text{tot}}$.
Based on similar arguments, an increase in $m_a$ would lead to the condition $m^2_{\text{eff}} = m_a^2$ being satisfied sooner at the far resonance, which leads to a decrease in $r_>$, and an eventual increase in $P_{\text{tot}}$.

The dependence of the locations of the near and far resonance points and the adibaticities and the conversion probabilities on the parameters $n_0$ and $m_a$ are qualitatively indicated in Table~\ref{table:location}.
\begin{table}[t]
	\centering
	\renewcommand{\arraystretch}{1.4}
	\begin{tabular}{|c|c|c|c|c|}\hline
      & increase in $n_0$& decrease in $n_0$&increase in $m_a$& decrease in $m_a$\\
     \hline
     \multirow{2}{*}{near-resonance} & $r_<$ decreases & $r_<$ increases & $r_<$ unchanged & $r_<$ unchanged\\
    & $\ad (r_<)$ decreases & $\ad (r_<)$ increases & $\ad (r_<)$  unchanged & $\ad (r_<)$  unchanged \\
     \hline
      \multirow{2}{*}{far-resonance } & $r_>$ increases & $r_>$ decreases & $r_>$ decreases & $r_>$ increases\\
      & $\ad (r_>)$ decreases & $\ad (r_>)$ increases & $\ad (r_>)$ increases & $\ad (r_>)$ decreases \\
	\hline
	\end{tabular}
	\caption{Qualitative behavior of the location and the adiabaticity parameter at the near and far-resonance points with respect to changes in $n_0$ and $m_a$.}
	\label{table:location}
\end{table}
Quantitatively, the adiabaticities at the near and far resonance points can be expressed as
\begin{equation}
	\ad (r_<) \approx 2 \, g_{a\gamma}^2 \, \frac{1}{\omega_\gamma} \left( \frac{r_<}{r_0} \right) \frac{r_0}{3 \,c_B}\;,
	\quad \quad \ad(r_>) \approx 2 \, g_{a\gamma}^2\, \omega_\gamma \left( \frac{r_>}{r_0} \right)^{-2} \frac{B_0^2}{n_0}\, \frac{r_0}{3 \, c_n} \;.
\end{equation}
Note that, $r_<$ and $r_>$ themselves is a function of $B_0$, $n_0$ and $m_a$.
Using the approximate dependence of $r_<$ and $r_>$ as indicated in Eq. (\ref{eq:rlessapprox}) and (\ref{eq:rgtrapprox}), we find that
\begin{equation}
	\ad (r_<) \approx 2 \,g_{a\gamma}^2\, \frac{1}{\omega_\gamma} \left( \frac{c_B \, \omega_\gamma^2 \, B_0^2}{c_n \, n_0} \right)^{1/3} \frac{r_0}{3 \, c_B} \; , \qquad	\ad (r_>) \approx 2\, g_{a\gamma}^2\, \omega_\gamma \left( \frac{c_n \, n_0}{m_a^2}\right)^{-\frac{2}{3}} \frac{B_0^2}{n_0} \, \frac{r_0}{3 \, c_n} \; .
	\label{eq:gammalessgtr}
\end{equation}
For typical values of parameters for a magnetar, taking $r_0 =10$ km, $B_0 = 1.35 \times 10^{14}$ G, and $n_0 = 10^{15}$ cm$^{-3}$, and further taking $m_a = 10^{-16}$ eV, for a photon of energy $\omega_\gamma =0.83$ keV, the values of $	\ad (r_<)$ and 	$\ad (r_>)$ are
\begin{equation}
	\ad (r_<) \approx 1.4 \; , \quad \quad \ad (r_>) \approx 5.1 \times 10^{-12} \; .
\end{equation}
Note from Eq.~(\ref{eq:gammalessgtr}), that the adiabaticity parameter at the near-resonance behaves as $	\ad (r_<)  \propto n_0^{-1/3}$, whereas the adiabaticity parameter at the far-resonance behaves as $	\ad (r_>)  \propto n_0^{-5/3} m_a^{4/3}$.
Therefore, the larger ratio of $\ad (r_>) /\ad (r_<) $ can be obtained for the smaller value of $n_0$ and larger values of $m_a$. For all ALPs with $m_a \lesssim 10^{-11}$ eV, even taking the smallest possible value of $n_0 = 10^{12}$ cm$^{-3}$, we find
\begin{equation}
	\ad (r_>) /\ad (r_<) \Big|_{\max} \; \approx \; 0.16 \;.
\end{equation}
Therefore, the contribution from the far-resonance can be safely ignored for estimating the transition probability near the magnetar.
In the limit of $	\ad (r_<)  \gg \ad (r_>)$, we have
\begin{equation}
	\mathbb{P} (\gamma_\parallel \to a) \approx P (r_<) \approx \left( 1- e^{-\pi \ad  (r_<) /2} \right) \; . \label{eq:singresapprox}
\end{equation}
The net $\gamma \to a$ conversion probability will include an extra factor of $1/2$ in front to account for $\gamma_\perp$ not coupling to ALPs, thus giving us
\begin{equation}
	\mathbb{P} (\gamma \to a) \approx \frac{1}{2} \left( 1- e^{-\pi \ad  (r_<) /2} \right) \; . \label{eq:singresapprox2}
\end{equation}

In reality, we expect that fluctuation and random orientations of the transverse magnetic fields will lead to many resonances being experienced by the photons. The formalism in section~\ref{sec:multres} will enable us to treat such a scenario. Due to the absence of detailed knowledge of the fluctuations involved, we estimate the lower limit of the conversion probability in the magnetar neighborhood using Eq.~(\ref{eq:ptot}) to be
\begin{equation}
P_{MN}{(\min)} =  \frac{1}{3} \left( 1 - e^{-3 \pi \ad (r_<) /4} \right) \lesssim \frac{1}{3} \left( 1 - e^{-\frac{3 \pi}{4} \ad_{\text{tot}}} \right)  \;.
	\label{eq:ptotagain}
\end{equation}
Here we have used
\begin{equation}
	\ad (r_<) \; \lesssim \, \sum \ad_i  \; = \, \ad_{\text{tot}}   \; .
\end{equation}
Note that this probability $P_{MN}{(\min)}$ is also less than that obtained by neglecting fluctuations in Eq.~(\ref{eq:singresapprox2}). Therefore, $P_{MN}{(\min)}$ is the conservative lower bound on the conversion probability in the magnetar neighborhood, with or without fluctuations.

\begin{figure}[t]
	\centering
	\includegraphics[width=.47 \textwidth]{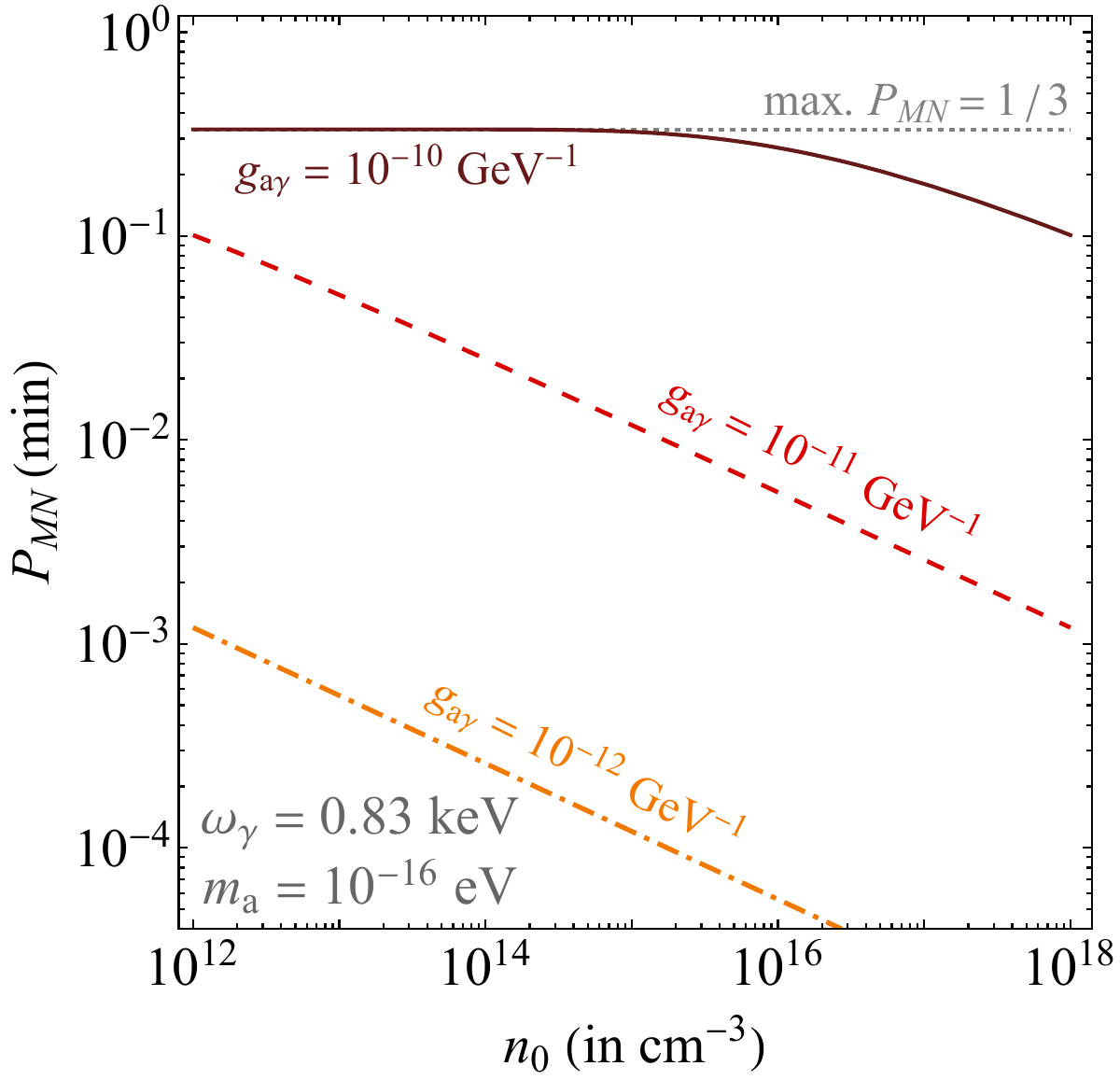}
	\hspace{5pt}
	\includegraphics[width=.48 \textwidth]{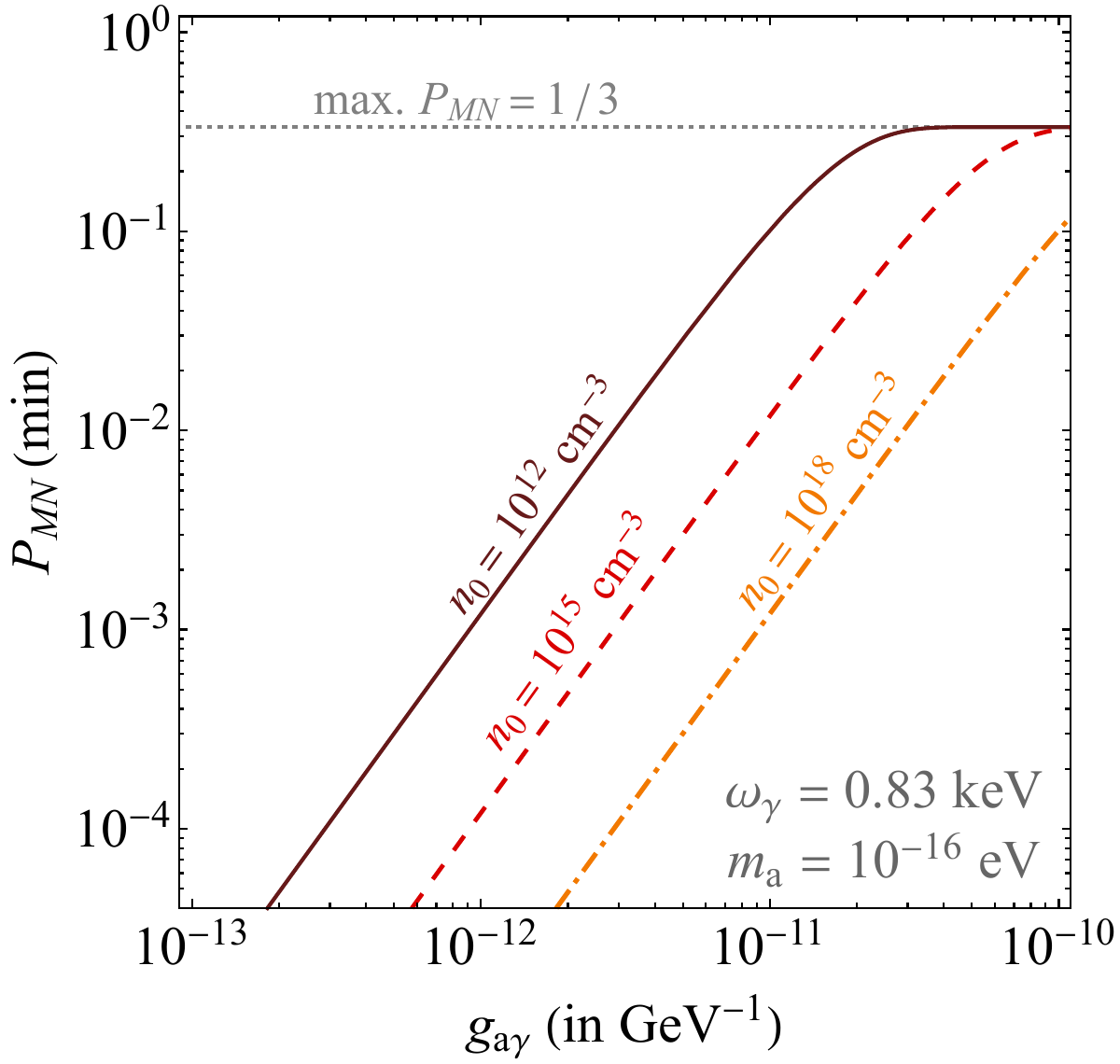}
	\caption{Resonance probability behavior near the magnetar. In the left panel we plot $P_{MN}{(\min)}$ vs. $n_0$, for different $g_{a\gamma}$ values. In the right panel, we plot $P_{MN}{(\min)}$ vs. $g_{a\gamma}$, for different possible $n_0$ values. The maximum value of $1/3$ is denoted by the gray dotted line. We take $\omega_\gamma = 0.83$ keV and $m_a =10^{-16}$ eV.}
	\label{fig:Ptot}
\end{figure}

In Fig.\,\ref{fig:Ptot}, we plot $P_{MN}{(\min)}$ as estimated in Eq.~(\ref{eq:ptotagain}), for the $\gamma \to a$ process near the magnetar. In the left panel, we plot the behavior of $P_{MN}{(\min)}$ as a function of $n_0$, for different values of $g_{a\gamma}$. As expected, $P_{MN}{(\min)}$ decreases with increasing values of $n_0$. Also, understandably, we observe that $P_{MN}{(\min)}$ decreases with decrease in the coupling $g_{a\gamma}$. In the right panel, we focus further upon this behavior and show the behavior of $P_{MN}{(\min)}$  as a function of $g_{a\gamma}$ for different values of $n_0$. At larger values of $g_{a\gamma}$, as well as at smaller values of $n_0$, the conversion probability asymptotically increases to $1/3$. 
This maximum conversion probability is denoted by the gray dotted lines.

\subsection{The Conversion Probability $P_{ISM}$ in the Interstellar Medium}
\label{sec:pISM}
The ALPs generated in the magnetar neighborhood will pass through the ISM where they may undergo further oscillations and partially convert back into photons. In this region, the magnetic field and the electron density can be approximated to be~\cite{Cordes:2002wz,Gaensler:2008ec,Jansson_2012,Miller:2013nza,planckintermediate}
\begin{equation}
	B_\text{ISM} \sim O( \text{few})\; \mu \text{G}\;,\qquad n_{\text{ISM}} \sim O(\text{few}\times 10^{-2}) \text{ cm}^{-3}\;.
\end{equation}
We take $B_\text{ISM} = 2 \; \mu \text{G}$ and $n_{\text{ISM}} = 2 \times 10^{-2} \text{ cm}^{-3}$ for our ballpark estimates.
For soft X-ray ($\sim\,$keV) photons, the magnitudes of the elements of the Hamiltonian may be expressed as
\begin{align}
	|\Delta_{a\gamma}| \approx & \; 1.52 \times 10^{-1} \left( \frac{g_{a\gamma}}{10^{-10} \text{ GeV}^{-1}} \right) \left( \frac{ B_{\text{ISM}} }{1 \mu\text{G}} \right) \text{ kpc}^{-1}\; , \\
	|\Delta_{pl}| \approx & \; 1.08 \left( \frac{ n_e }{ 10^{-2} \text{ cm}^{-3} } \right) \left( \frac{ 1\text{ kev} }{ \omega_\gamma } \right) \text{ kpc}^{-1} \; ,\\
	|\Delta_a| \approx & \; 7.82 \times 10^{-2} \left( \frac{ m_a }{ 10^{-12} \text{ eV} } \right)^2 \left( \frac{ 1\text{ keV} }{ \omega_\gamma } \right) \text{ kpc}^{-1} \; .
\end{align}
It may be observed that the dominant contribution in the oscillation frequency $\Delta_{\text{osc}}$ would be from $\Delta_{pl}$, when the ALP mass $m_a \lesssim 10^{-12}$ eV. The oscillation length scale associated with this term is
\begin{equation}
	\lambda_{pl} \equiv  \frac{2 \pi }{|\Delta_{pl}| } \approx 5.82 \times  \left( \frac{ 10^{-2} \text{ cm}^{-3}  }{n_e } \right) \left( \frac{ \omega_\gamma }{ 1\text{ keV}  } \right) \text{ kpc}\;.
\end{equation}
This gives, for $n_{\text{ISM}} = 2 \times 10^{-2} \text{ cc}^{-1}$ and for $\omega_\gamma = (0.71 - 0.98)$ keV, an oscillation length of
\begin{equation}
	\lambda_{pl} \approx (2.06 -  2.84) \text{ kpc}.
\end{equation}
Note that the distance of the Magnetar candidate under consideration is $L = 9$ kpc.
This means that the distance traveled alone doesn't allow an averaging out of the probability, as the dominant oscillation length scale $\lambda_{pl}$ is comparable to distance of $L = 9$ kpc. However, due to the large energy band $\omega_\gamma = (0.71 - 0.98)$, the range of corresponding $\lambda_{pl}$ values is quite large and averaging over oscillation probabilities needs to be performed. The average oscillation probability is given by
\begin{align}
	P_{\text{ISM}} (a \to \gamma) \, = \; & \Biggr( \int_{E_{\min}}^{E_{\max}}     \left[\frac{4 \Delta_{a\gamma}^2 }{ \Delta_{\text{osc}}^2} \sin^2\left( \frac{\Delta_{\text{osc}}\, L}{2}\right) \frac{d\phi}{dE}\,  \right] dE  \Biggr) \Bigg/ \Biggr( \int_{E_{\min}}^{E_{\max}} \frac{d\phi}{dE}\, dE  \Biggr) \;,
	\label{eq:PavgEnergy}
\end{align}
where $d\phi/dE$ is the differential flux in the energy band. Approximating this differential flux to be uniform over the energy band, and averaging over the oscillating term $ \sin^2\left(\Delta_{\text{osc}}\, L/\, 2\right)$ we get
\begin{equation}
	P_{\text{ISM}} (a \to \gamma) \approx \, 2  \, \frac{\Delta_{a\gamma}^2}{\langle \Delta_{\text{osc}}^2\rangle}\; , \label{eq:pavgfinal}
\end{equation}
where we take $\langle \Delta_{\text{osc}}^2\rangle$ to be the value of $\Delta_{\text{osc}}^2$ at the average energy in the band.
Note that we have taken a constant potential (constant ISM magnetic field and electron density) approximation for the sake of simplification, while in reality we expect a varying electron density. 

\begin{figure}[t]
	\centering
	\includegraphics[width=.6 \textwidth]{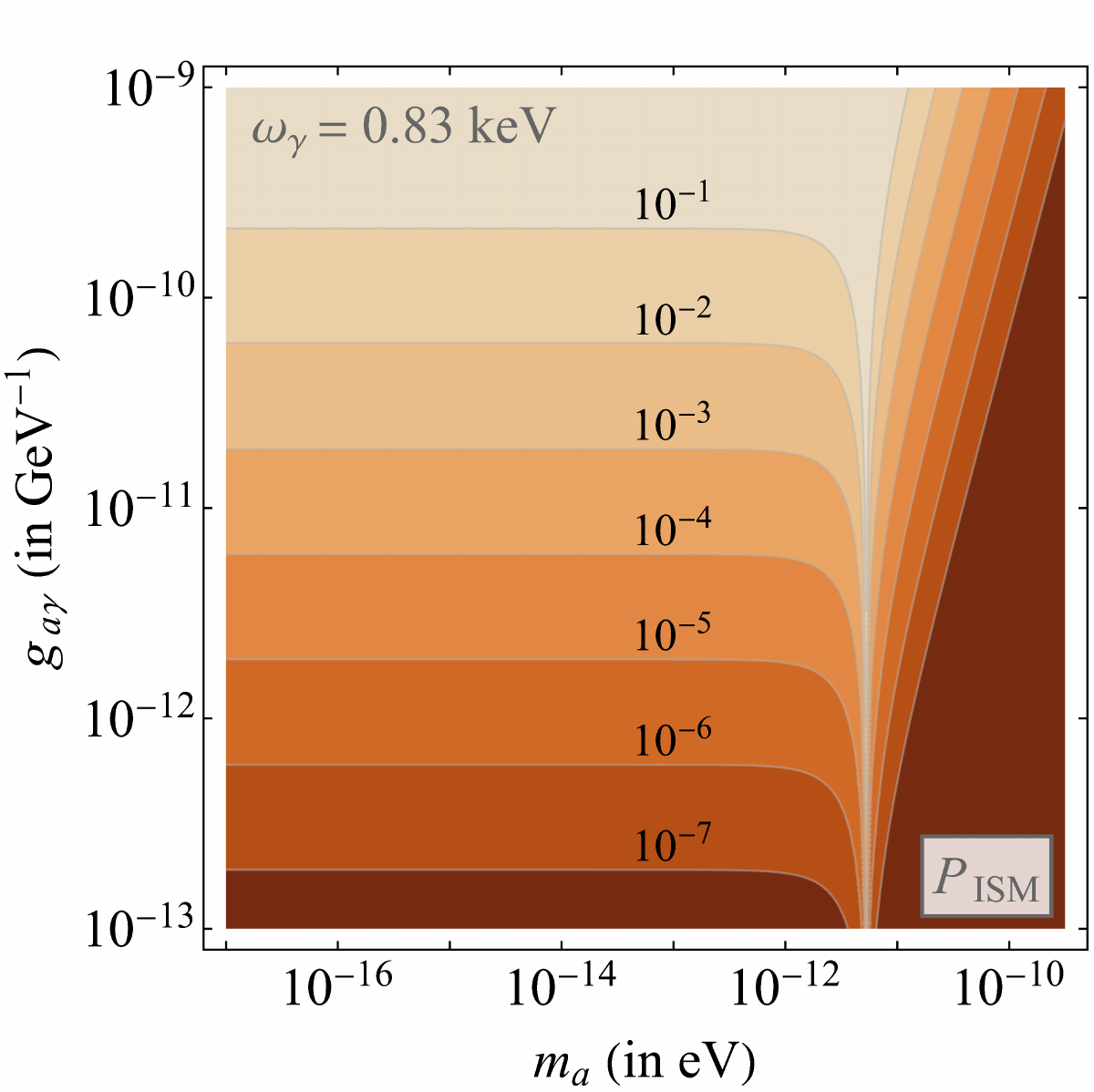}
	\caption{Conversion probability $P(a \to \gamma)$ in the ISM, using benchmark estimates of electron density $n_{\text{ISM}} = 2 \times 10^{-2}$ cc$^{-1}$ and magnetic field $B_\text{ISM} = 2 \; \mu \text{G}$, for a photon of energy  $\omega_\gamma = 0.83$ keV. The kink in the plot at $\sim (10^{-12} - 10^{-11})$ eV is due to the resonance with the effective mass of the photon in the ISM, which is dependent upon the benchmark parameter values chosen.}
	\label{fig:PISM}
\end{figure}

The conversion probability in the interstellar medium is shown in Fig.\,\ref{fig:PISM}. Note the sharp increase of the conversion probability $P_{\text{ISM}}$ at $m_a \sim O(10^{-12} - 10^{-11})$ eV. This is caused by the resonance due to the ALP mass matching the photon effective mass in the ISM. However, this sharp feature is only due to the use of the average electron density in the ISM. In reality, due to varying electron density along the line of sight of the photon, this feature will be more spread out and less prominent. Due to the uncertainties involved with the location and the contribution of the ISM resonance peak, our results in the range $m_a \sim (10^{-11}-10^{-12})$ eV have to be interpreted with caution. We therefore focus on the low-mass regime ($m_a \lesssim 10^{-12}$~eV) and the high-mass regime ($m_a \gtrsim 10^{-11}$ eV).

In the low-mass ($m_a \lesssim 10^{-12}$ eV) regime, the average $a \to \gamma$ conversion probability in the ISM can be approximated as
\begin{equation}
 P_{\text{ISM}}(a \to \gamma) \approx  \, 2  \, \Delta_{a\gamma}^2\, /\, \Delta_{pl}^2  \; \propto \; g_{a\gamma}^2 \;.
\end{equation}
This explains the horizontal contours on the left side of the plot, as the probability in this region only depends on the ALP-photon coupling $g_{a\gamma}$. On the other hand, in the high-mass $(m_a \gtrsim 10^{-11})$ regime , the conversion probability behaves as
\begin{equation}
P_{\text{ISM}} (a \to \gamma)  \approx  \, 2  \, \Delta_{a\gamma}^2\, /\, \Delta_{a}^2 \; \propto \; g_{a\gamma}^2 / m_a^2 \;,
\end{equation}
leading to a dependence on both $g_{a\gamma }$ and $m_a$. We observe this dependence on both the parameters on the right side of the plot, where, with increasing $m_a$ the conversion probability drops sharply.

\section{\label{sec:result}Constraints on the Photon-ALP Coupling}
We now obtain the constraints on $g_{a\gamma}$ by combining the results from sections \ref{sec:pMagnetar} and \ref{sec:pISM}. 
From Fig.\,\ref{fig:resonancecontour}, we observe that for our parameter ranges (in particular, for the range of possible $n_0$ values), resonances are inevitable for all ALPs with $m_a \lesssim 10^{-11}$ eV. A lower bound on the $\gamma \to a$ conversion probability $P_{MN}$ in the magnetar neighborhood has been estimated in section \ref{sec:pMagnetar}.
This conversion probability has a maximum possible value of $1/3$, where all three flavors are present in equal measure.
\begin{figure}[t]
	\centering
	\includegraphics[width=.6 \textwidth]{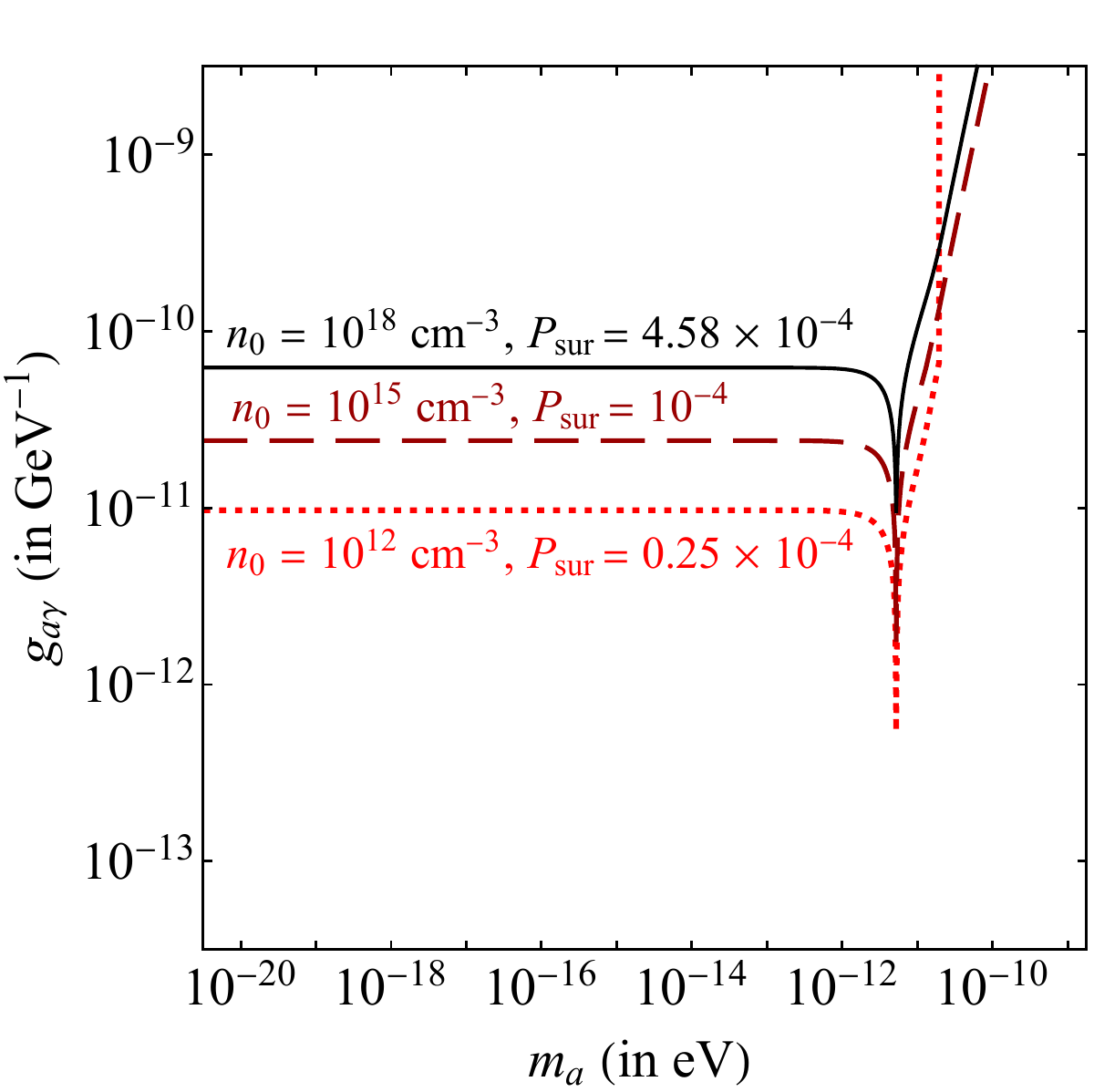}
	\caption{Upper bounds on $g_{a\gamma}$ as functions of $m_a$, for three benchmark values $n_0$ and $P_{\text{sur}}$, corresponding to the most optimistic, the most conservative and an intermediate scenario. The abrupt change in the slope of the red dotted line at $m_a \approx 10^{-11}$ eV is due the resonance condition not being fulfilled near the magnetar.}
	\label{fig:bound0}
\end{figure}

From section \ref{sec:pISM}, we obtain the probability of ALP to photon conversion in the ISM and its dependence on the ALP mass $m_a$ as well as the ALP-photon coupling $g_{a\gamma}$. This probability $P_{ISM}$ is obtained as $P_{\text{avg}} (a \to \gamma)$ given in Eq.~(\ref{eq:pavgfinal}).

To obtain bounds on ALP-photon conversion and therefore on the coupling $g_{a\gamma}$, we impose the condition that  the fraction of photons that may escape via the $\gamma \to a\to \gamma$ process must be less than the total fraction of photons that survived the journey from an obscured magnetar. Thus, the photon survival probability $P_{\text{sur}} (E_{\text{bin}}) \equiv F_{\text{obs}} (E_{\text{bin}})/F_{0} (E_{\text{bin}})$ must be larger than $P(\gamma \to a\to \gamma)$, i.e.,
\begin{equation}
	P_{\text{sur}}  (E_{\text{bin}}) \gtrsim \; P_{MN} \, P_{ISM} \gtrsim \; P_{MN}{(\min)}\, P_{ISM} \; .
\end{equation}
The $\gamma \to a$ conversion probability near the magnetar depends on the value of $n_0$, an increase in $n_0$ would lead to a decrease in the total conversion probability $P_{MN}$ near the magnetar.
The strongest bound on $g_{a\gamma}$ is therefore obtained for the smallest possible value of $P_{\text{sur}} $ and the smallest value of $n_0$. Whereas the most conservative bound would be obtained for the largest possible value of $P_{\text{sur}} $ and the largest value of $n_0$.

\begin{figure}[t]
	\centering
	\includegraphics[width= \textwidth]{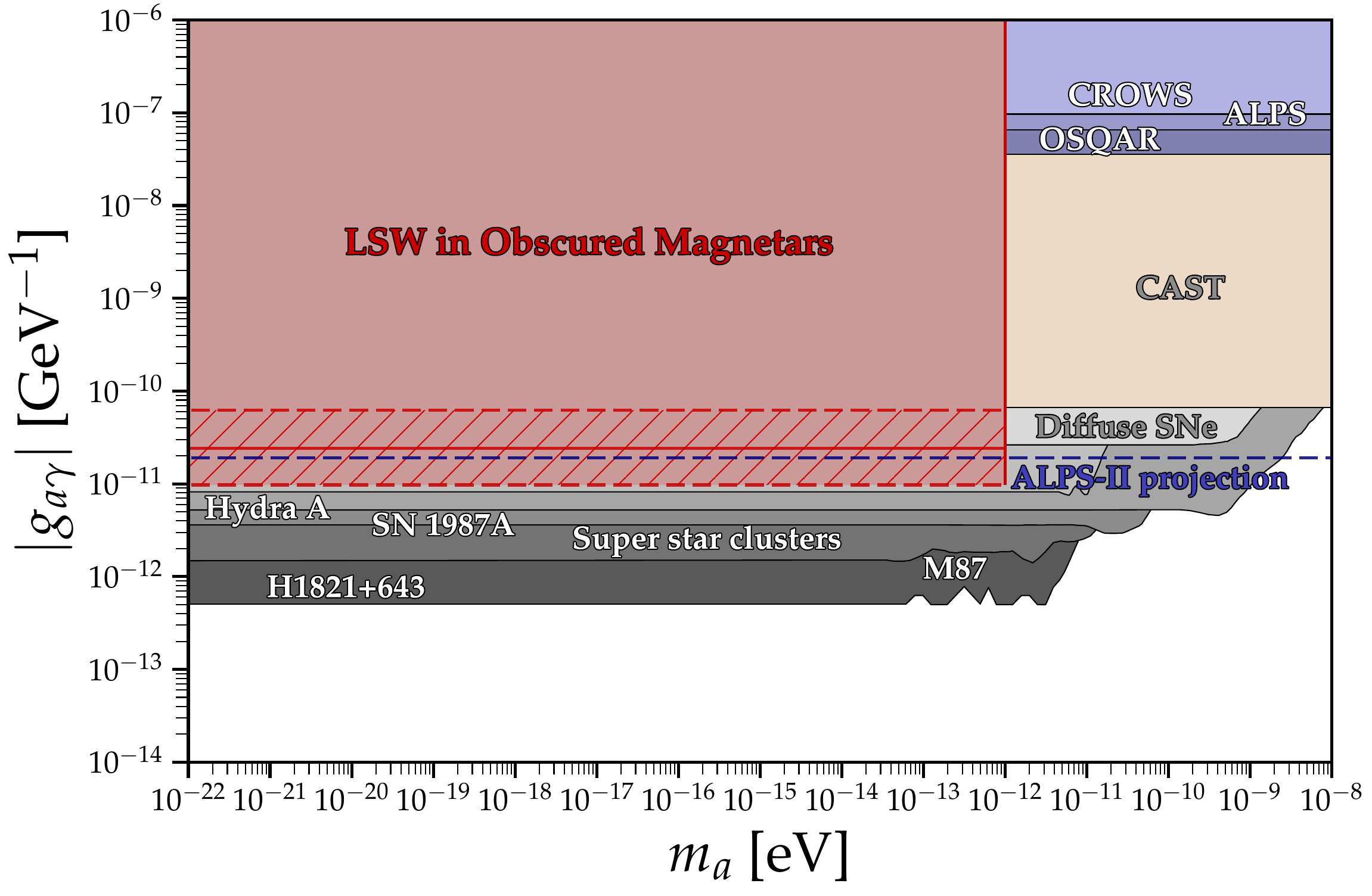}
	\caption{{Constraints on the ALP-photon coupling $g_{a\gamma}$ through applying the LSW technique on the obscured magnetar PSR J1622-4950. The constraint corresponding to the intermediate parameter values ($n_0 =10^{15}$ cm$^{-3}$, $P_{\text{sur}} = 10^{-4}$) is shown as a deep red line, whereas the light red region corresponds to the whole range of possible constraints on $g_{a\gamma}$. The dashed lines correspond to the most conservative and the most optimistic scenarios. The hatched region between them corresponds to the uncertainty, which arises mainly from the large range of $n_0$ values considered. The constraints obtained in this work are compared against existing and future lab-based LSW~\cite{Ehret:2010mh,Betz:2013dza,Ballou:2014myz,Ortiz:2020tgs}, CAST~\cite{CAST:2004gzq,CAST:2017uph} and astrophysical constraints~\cite{Wouters:2013hua,Marsh:2017yvc,Meyer:2020vzy,Dessert:2020lil,Reynes:2021bpe,Hoof:2022xbe}.}}
	\label{fig:bound1}
\end{figure}

In order to get the strongest bound, we choose the bin $\omega_\gamma = (0.71 - 0.98)$ keV, where the ratio between observed flux $F_{\text{obs}}$ and intrinsic flux $F_0$
is the lowest. In this bin $P_{\text{sur}} = (0.25 - 4.58) \times 10^{-4}$ for the magnetar PSR J1622-4950. 
In Fig.\,\ref{fig:bound0}, we show the bounds for the two extreme scenarios as well as an intermediate benchmark scenario:
\begin{itemize}
	\item \textit{the conservative scenario} with the highest values of $n_0$ as well as $P_{\text{sur}}$ ($n_0 =10^{18}$ cm$^{-3}$, $P_{\text{sur}} = 4.58 \times 10^{-4}$), denoted by the black line in Fig.\,\ref{fig:bound0}.
	\item \textit{the optimistic scenario} with the lowest values of $n_0$ as well as $P_{\text{sur}}$ ($n_0 =10^{12}$ cm$^{-3}$, $P_{\text{sur}} = 0.25 \times 10^{-4}$), denoted by the red dotted line in the figure.
	\item \textit{the benchmark scenario} with an intermediate set of parameters ($n_0 =10^{15}$ cm$^{-3}$, $P_{\text{sur}} = 10^{-4}$), denoted by the maroon dashed line in the figure.
\end{itemize}

 For $m_a \lesssim 10^{-12}$ eV, the LSW technique applied in the context of obscured magnetars constrains $g_{a\gamma} \lesssim (10^{-10} - 10^{-11})$ GeV$^{-1}$. The kink at $m_a \sim 10^{-11}$ eV is due to a possible resonance in ISM, as discussed in Section~\ref{sec:pISM}. However, any actual effect of a resonance in ISM, if at all present, will be sensitive to the variations in $n_{\text{ISM}}$ and $B_{\text{ISM}}$, resulting in the smearing out of the features.
For the high-mass region of $m_a \gtrsim 10^{-11}$ eV, resonances may not happen even in the magnetar neighborhood. In such cases, our approximations would yield no photons escaping through the $\gamma \to a\to \gamma$ process, and therefore, no constraints on $g_{a\gamma}$ would be obtained. This feature can be observed in the abrupt change in the slope of the red dotted line corresponding to the lowest $n_0$ value of $n_0 =10^{12}$ cm$^{-3}$ in Fig.\,\ref{fig:bound0}.

In Fig.\,\ref{fig:bound1}, the constraints on $g_{a\gamma}$ obtained in this work are compared against already existing bounds in the literature. Our constraints are shown as the light red region. The uncertainty at the lower end of this region, showed with hatching, is mainly due to the large range of $n_0$ values considered.
The regions with different shades of purple correspond to the exclusion limits given by the laboratory based LSW experiments (ALPS~\cite{Ehret:2010mh}, CROWS~\cite{Betz:2013dza}, and OSQAR~\cite{Ballou:2014myz}). It may be seen that the bound obtained in this work is $\sim \! O(10^3)$ times stronger than the lab-based LSW bounds. 
The blue dashed line is the projected future constraint from the ALPS-II experiment~\cite{Ortiz:2020tgs}.
Note that our method enables us to obtain results competitive with the ALPS-II experiment with already existing data.
The light yellow region shows the constraint from the CAST experiment~\cite{CAST:2004gzq,CAST:2017uph}, and the different shades of gray denote the astrophysical constraints~\cite{Wouters:2013hua,Marsh:2017yvc,Meyer:2020vzy,Dessert:2020lil,Reynes:2021bpe,Hoof:2022xbe}. The bound on $g_{a\gamma}$ is complementary to existing astrophysical bounds.

Note further, that we only plot for the low mass regime of $m_a \lesssim 10^{-12}$ eV.
The high mass regime is not considered, since here the resonance condition near a magnetar may not be fulfilled, or we may obtain a resonance in ISM, confounding our analysis.

\section{\label{sec:conclusion}Concluding Remarks}

In this work, we explore whether the light shining through wall (LSW) technique used to hunt for ALPs in laboratory can be applied in an astrophysical setting.
We find that obscured magnetars are excellent candidates, due to their high luminosity, large magnetic fields, and the presence of an astrophysical wall in the form of a nebula. In the presence of ALP-photon oscillations, the number of photons from the magnetar that reach us would be larger than the number of photons that we would observe otherwise.

The fraction of photons that we are able to observe can be estimated by comparing the intrinsic luminosity against the observed number of photons in any energy band. This total fraction of observed photons must be inevitably larger than the fraction that may escape through the $\gamma \to a \to \gamma$ process.
To estimate the fraction of photon escaping through the latter process, we calculate the conversion probabilities in the magnetar neighborhood as well as in the interstellar medium. 
In the magnetar neighborhood, we investigate the effects of possible resonances encountered by the photon, and use a conservative analytic estimate to calculate the number of photons that may be converted into ALPs.
In the interstellar medium, we estimate the conversion probability by using the typical average values of the magnetic field and  the electron density.
This allows us to obtain the dependence of $P (\gamma \to a \to \gamma)$ on the ALP-photon coupling, and hence estimate an upper limit on $g_{a\gamma}$. 

As a proof of concept, we use the observation of the magnetar PSR J1622-4950 by the XMM-Newton X-ray telescope. In the energy bin of $(0.71 - 0.98)$ keV, the survival probability of photons is the smallest, and hence gives us the strongest bounds.
Our technique leads to a constraint of $g_{a\gamma} \lesssim (10^{-10} -10^{-11})$ GeV$^{-1}$ for $m_a \lesssim 10^{-12}$ eV.
This is $\sim O(10^3)$ better than existing laboratory based LSW bounds obtained thus far, it is also competitive with ALPS-II projections at the low mass regime. 
The obtained bound is weaker than some of the existing astrophysical constraints, but this technique is capable of giving a stronger bound if a more obscured magnetar, with even lower survival probability of photons, is observed in future.

A more detailed treatment with observation of multiple magnetars, as well as a better understanding of the spatial behavior of the electron density and magnetic field in the magnetar neighborhood can make the bounds more robust. Below, we outline some of the possible ways by which the bounds on $g_{a\gamma}$ can be strengthened:
\begin{itemize}
	\item More obscured (fainter, due to attenuation of light in the nebula) magnetars may lead to stronger constraints on $g_{a\gamma}$. Magnetars with denser/ larger nebulas in the foreground would be good candidates to search for.
	\item A more dedicated (longer) observations of potential obscured magnetar candidates may lead to better bounds, as this may give lower values of photon survival probability $P_{\text{sur}}$.
	\item As discussed in this work, a magnetar with lower value of electron density ($n_0$) would lead to a stronger constraint.
	\item Non-observation of any flux at a particular energy bin may also lead to stronger constraints, since this would imply that the number of photons reaching us in that energy bin is lower than the sensitivity of the detector.
\end{itemize}
For example, if we can establish a survival probability of $P_{\text{sur}} \approx 10^{-8} $ at $\sim 1$ keV bin, and find a magnetar with $n_0 \sim 10^{13}$ cm$^{-3}$, $B_0 \sim 10^{15}$ G, the constraint could become as strong as $g_{a\gamma} \lesssim 10^{-12} \text{ GeV}^{-1}$. Such a value of $P_{\text{sur}} \sim 10^{-8}$ is possible to obtain with an observation of $\sim 10$ days and the effective photo-detector area of $\sim \text{few} \times10^5$~cm$^2$, when the intrinsic flux in the observed bin is $F_0 \approx \text{few} \times 10^{-13} \text{ erg cm}^{-2} \text{ s}^{-1}\, $.

The concept of applying the light shining through wall (LSW) technique in an astrophysical setting has been implemented in a novel manner in this work. Furthermore, to account for our less-than-perfect understanding of the magnetar and the nebula, we have taken conservative estimates at each step. In future, better understanding and dedicated searches may enable us to put even stronger bounds  on $g_{a\gamma}$ using the idea outlined in this work.

\section*{Appendices}

\appendix

\section{The Effects of ALP-Photon Conversion in the Nebula}
\label{sec:contribution}

\begin{figure}[t]
	\centering
	\includegraphics[width=\textwidth]{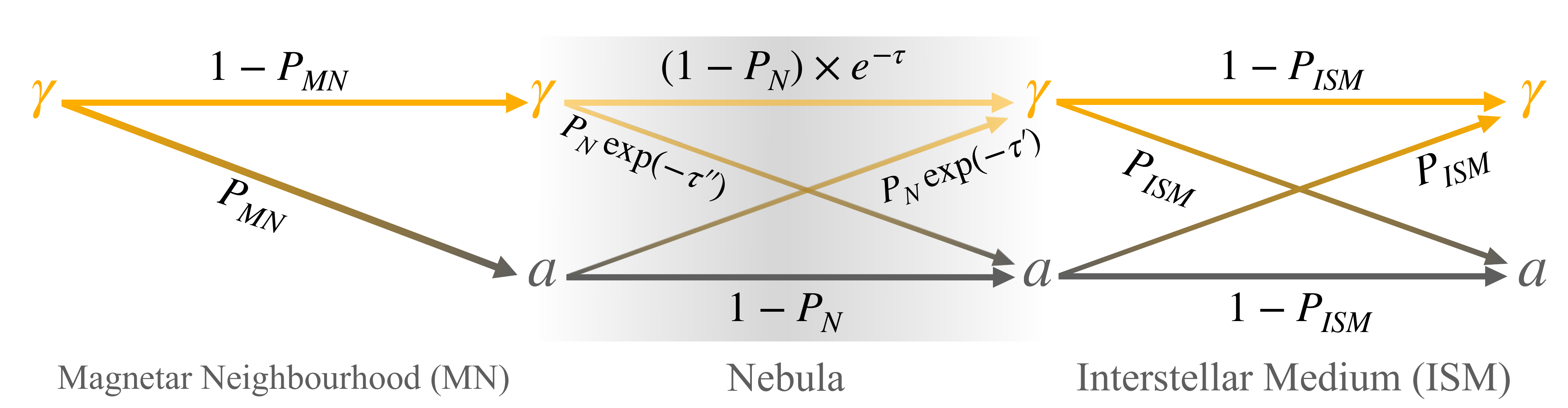}
	\caption{Schematic diagram representing the conversion and survival probabilities in different regions as a photon propagates from the obscured magnetar to an observer. All possible states of the system are illustrated at different times during propagation. Here, $P_{MN}$ denotes the $\gamma \to a$ conversion probability in the magnetar neighborhood (MN), $P_{N}$ denotes the conversion probability in the nebula (N), and $P_{ISM}$ denotes the $a\to \gamma$ conversion probability in the interstellar medium (ISM). The attenuation of photons in the nebula is represented by the opacity $\tau$.  The quantities $\tau^\prime$ and $\tau^{\prime\prime}$ denote the effective opacities for photons that convert from or into ALPs somewhere in the nebula. The scenario with $P_N \approx 0$ would correspond to Fig.\,\ref{fig:LSWastro}.}
	\label{fig:LSWastroAppendix}
\end{figure}
In the main text we approximated the propagation through nebula such that the photons are absorbed and ALPs pass completely undisturbed. However, as we know, there would be some ALP-photon conversion which will change this picture somewhat. This is indicated schematically in Fig.\,\ref{fig:LSWastroAppendix}, with the introduction of the modified diagrams for the nebula.
Here $P_N$ is the net conversion probability in the nebula, $\tau$ is the opacity experienced by a photon as it propagates through the nebula, and the quantities $\tau^\prime$ and $\tau^{\prime\prime}$ denote the effective opacity experienced by ALPs that convert to photons and photons that convert to ALPs, respectively, somewhere in the nebula.

The flux that may escape via oscillating into ALPs, through the $\gamma \to a \to \gamma$ LSW mechanism (process C in section~\ref{sec:threeprocesses}) can be expressed in detail as
\begin{align}
	F^{\text{LSW}} (E_\text{bin}) \approx & \,  F_{0} (E_\text{bin}) \, P_{MN} \, (1-P_N) \, P_{ISM}\;. \nonumber \\
+	& \, F_{0} (E_\text{bin}) \, P_{MN} \, P_{N} \, \exp[-\tau^{\prime} ] \, (1- P_{ISM} ) \nonumber \\
+	& \, F_{0} (E_\text{bin}) \, (1-P_{MN} ) \, P_{N} \, \exp[-\tau^{\prime\prime} ] \, P_{ISM}\;.
	\label{eq:escape}
\end{align}
Therefore, 
\begin{align}
	F^{\text{LSW}} (E_\text{bin}) \gtrsim & \,  F_{0} (E_\text{bin}) \, P_{MN} \, (1-P_N) \, P_{ISM}\;.	\label{eq:escapeapprox}
\end{align}
Typically, away from the magnetar, we would expect that $P_N$ would be very small, i.e. $P_N \approx 0$ and the scenario shown in Fig.\,\ref{fig:LSWastro} would be reproduced.
Even in extreme case, the maximum value that $P_N$ can take would be at the most $1/3$, which will occur when both the photon polarizations and the ALP mix efficiently in the nebula. Therefore $(1-P_N) \ge 2/3$ at all times. This can at most lead to a $\sim20\%$ relaxation of our constraints.

\section{Photon-ALP Conversion Probability for Multiple Domains}
\label{sec:appendix}
In this appendix, we calculate the photon-ALP conversion probability during their propagation through multiple domains of magnetic field and electron density.
To simplify the problem, we assume that the orientation of the magnetic field is random.
We generalize the results derived in~\cite{Grossman:2002by} to the case where the transition probability in each domain is arbitrary. The problem is treated in the full ``three-flavor'' scenario.

For photon polarizations, we use the basis defined in terms of their components at the start of the first domain, as
\begin{equation}
	 | \gamma_1 \rangle \equiv  | \gamma_\parallel (0) \rangle \; , \qquad  | \gamma_2 \rangle \equiv | \gamma_{\perp} (0) \rangle \;.
\end{equation} 
Let the initial intensities of $\gamma_1$, $\gamma_2$ and $a$ be $I_{\gamma_1} (0)$, $I_{\gamma_2} (0)$ and $I_a (0)$. 

In the $n$-th domain, let the transverse component of the magnetic field be rotated by an angle $\theta_n$ compared to the first domain. The photon polarizations, parallel and perpendicular to the transverse magnetic field in this domain can be expressed as
\begin{align}
	| \gamma_\parallel (n) \rangle = \cos \theta_n | \gamma_1 \rangle  + \sin \theta_n | \gamma_2 \rangle \; , \qquad 	| \gamma_\perp (n) \rangle = -\sin \theta_n | \gamma_1 \rangle  + \cos \theta_n | \gamma_2 \rangle \;.
\end{align}
In terms of the $( | \gamma_1 \rangle, \; | \gamma_2 \rangle )$ basis, we then have
\begin{equation}
	| \gamma_1 \rangle = \cos \theta_n \, | \gamma_\parallel (n)\rangle- \sin \theta_n \, | \gamma_\perp (n) \rangle \; , \qquad |	\gamma_2 \rangle= \sin \theta_n \, | \gamma_\parallel (n) \rangle + \cos \theta_n \, | \gamma_\perp (n) \rangle  \; .
\end{equation}
The Photon-ALP transition in the $n$-th domain involves only $| \gamma_\parallel (n) \rangle$ and the ALP state $|a \rangle$.
Let this transition probability be $p_n$, then at the end of $n$-th domain the photon and ALP intensities will be
\begin{align}
	I_{\gamma_1} (n) \sim & \left(1- \cos^2 \theta_n \, p_{n} \right) I_{\gamma_1} (n-1) +\cos^2 \theta_n\, p_{n} \, I_a (n-1) \; ,\\[6pt]
	I_{\gamma_2} (n) \sim & \left(1- \sin^2 \theta_n \, p_{n} \right) I_{\gamma_2} (n-1) +\sin^2 \theta_n \, p_{n} \, I_a (n-1) \; ,\\[6pt]
	I_a (n) \sim & \; p_{n}\left[ \cos^2 \theta_n \, I_{\gamma_1} (n-1)+ \sin^2 \theta_n \, I_{\gamma_2} (n-1)\right]  + (1- p_{n}) \, I_a (n-1) \;.
\end{align}
Note that due to the random nature of $\theta_n$, the $\cos^2 \theta_n$ and $\sin^2 \theta_n$ may be replaced by their averaged value of $1/2$ when we are dealing with a large number of domains.
Then we get the ``averaged'' recursion relation
\begin{equation}
	\left(\begin{array}{c}
		I_{\gamma_1} (n) \\[5pt]
		I_{\gamma_2} (n) \\[5pt]
		I_a (n) \\
	\end{array}\right) =  
	\left(\begin{array}{ccc}
		1- \frac{1}{2}p_{n} & 	0& 	\frac{1}{2} p_{n} \\[5pt]
		0 & 	1- \frac{1}{2}p_{n} & 	\frac{1}{2} p_{n} \\[5pt]
		\frac{1}{2} p_{n}&	\frac{1}{2} p_{n}&1- p_{n}
	\end{array}\right)
	\left(\begin{array}{c}
		I_{\gamma_1} (n-1) \\[5pt]
		I_{\gamma_2} (n-1) \\[5pt]
		I_a (n-1) \\
	\end{array}\right)  \;.
\end{equation}
This recursion relation leads to the intensity at the end of $n$-th domain evolving as
\begin{equation}
	\left(\begin{array}{c}
	I_{\gamma_1} (n) \\[5pt]
	I_{\gamma_2} (n) \\[5pt]
	I_a (n) \\
\end{array}\right) =	\mathbb{M}_{3\times3}^{(n)} 	\left(\begin{array}{c}
		I_{\gamma_1} (0) \\[5pt]
		I_{\gamma_2} (0) \\[5pt]
		I_a (0) \\
	\end{array}\right) \; .
\end{equation}
Here,
\begin{equation}
	\mathbb{M}_{3\times3}^{(n)} = \left(\begin{array}{ccc}
 	m_{11}^{(n)} \; & \; m_{12}^{(n)} \; & \; m_{13}^{(n)} \\[5pt]
 	m_{12}^{(n)} \; & \; m_{11}^{(n)} \; & \; m_{13}^{(n)} \\[5pt]
 	m_{13}^{(n)} \; & \; m_{13}^{(n)} \; & \; m_{33}^{(n)} \\
	\end{array}\right) \; ,
\end{equation}
with the individual elements of the matrix given by
\begin{align}
	m_{11}^{(n)} = & \, \frac{1}{3} + \frac{1}{6} \prod_{k=1}^{n} \left( 1 - \frac{3}{2} p_k \right) +\frac{1}{2} \prod_{k=1}^{n} \left( 1 - \frac{1}{2} p_k \right) \; ,\label{eq:m11}\\
	m_{12}^{(n)} = & \, \frac{1}{3} + \frac{1}{6} \prod_{k=1}^{n} \left( 1 - \frac{3}{2} p_k \right) -\frac{1}{2} \prod_{k=1}^{n} \left( 1 - \frac{1}{2} p_k \right) \; ,\label{eq:m12}\\
	m_{13}^{(n)} = & \frac{1}{3} \left[1 - \prod_{k=1}^{n} \left( 1 - \frac{3}{2} p_k \right)  \right]\; ,\label{eq:m13}\\
	m_{33}^{(n)} = & \frac{1}{3} \left[1 + 2 \prod_{k=1}^{n} \left( 1 - \frac{3}{2} p_k \right)  \right]\; . \label{eq:m33}
\end{align}
The method of induction can be used to show that the Eqs.~(\ref{eq:m11}-\ref{eq:m33}) holds for any integer value of $n$.
For $n=1$, the above equations simplify to the expected expression
\begin{equation}
		\mathbb{M}_{3\times3}^{(1)} =  	\left(\begin{array}{ccc}
			1- \frac{1}{2}p_{1} & 	0& 	\frac{1}{2} p_{1} \\[5pt]
			0 & 	1- \frac{1}{2}p_{1} & 	\frac{1}{2} p_{1} \\[5pt]
			\frac{1}{2} p_{1}&	\frac{1}{2} p_{1}&1- p_{1}
		\end{array}\right) \; .
\end{equation}
Similarly, for the $n=2$ limit, we find
\begin{equation}
	\mathbb{M}_{3\times3}^{(2)} =  	\left(\begin{array}{ccc}
		1- \frac{1}{2}p_{2} & 	0& 	\frac{1}{2} p_{2} \\[5pt]
		0 & 	1- \frac{1}{2}p_{2} & 	\frac{1}{2} p_{2} \\[5pt]
		\frac{1}{2} p_{2}&	\frac{1}{2} p_{2}&1- p_{2}
	\end{array}\right) \cdot \left(\begin{array}{ccc}
		1- \frac{1}{2}p_{1} & 	0& 	\frac{1}{2} p_{1} \\[5pt]
		0 & 	1- \frac{1}{2}p_{1} & 	\frac{1}{2} p_{1} \\[5pt]
		\frac{1}{2} p_{1}&	\frac{1}{2} p_{1}&1- p_{1}
	\end{array}\right)  \; .
\end{equation}
Finally, for $n$ domains, it can be shown that:
\begin{equation}
		\mathbb{M}_{3\times3}^{(n)}  = 	\left(\begin{array}{ccc}
		1- \frac{1}{2}p_{n} & 	0& 	\frac{1}{2} p_{n} \\[5pt]
		0 & 	1- \frac{1}{2}p_{n} & 	\frac{1}{2} p_{n} \\[5pt]
		\frac{1}{2} p_{n}&	\frac{1}{2} p_{n}&1- p_{n} 	
	\end{array}\right) \cdot \mathbb{M}_{3\times3}^{(n-1)} 
\end{equation}
Therefore, the conversion probability after passing through $n$ domains, can be expressed as
\begin{equation}
	\mathbb{P} (\gamma \to a) = \frac{1}{3} \left[ 1-\prod_{k=1}^{n} \left(1 - \frac{3}{2} \, p_k \right)  \right] \;.
	\label{eq:appendix_conv_1}
\end{equation}
In the limit of large $n$, using the limiting value
\begin{equation}
	\prod_{k=1}^{n} \left( 1 - \frac{3}{2} p_k \right) \quad \overset{(n \to \infty)}{\longrightarrow} \quad \exp \left( - \frac{3}{2} \sum_{k=1}^{n}p_k \right) \; ,
\end{equation}
Eq.~(\ref{eq:appendix_conv_1}) can be approximated as
\begin{equation}
	\mathbb{P} (\gamma \to a) = \frac{1}{3} \left[ 1-\exp \left( - \frac{3}{2} \sum_{k=1}^{n}p_k \right)  \right] \; .
	\label{eq:appendix_conv_2}
\end{equation}
This is the conversion probability, as given in Eq.~(\ref{eq:ptot}). Note that, this generalized answer, in the limit of identical $p_k$'s, matches the conversion probability obtained in~\cite{Grossman:2002by}.

\acknowledgments{The authors acknowledge support from the Department of Atomic Energy (DAE), Government of India, under Project Identification No. RTI4002. DSC would like to thank G. Raffelt and S. Abbar for their inputs during initial discussions, S. Kulkarni and Nirmal Raj for their suggestions and comments, and S. Mukherjee for his help in navigating the astrophysical data.
DSC would also like to thank S. Witte, E. Fuchs and T. Bringmann for their comments.
Further, DSC thanks the Light Dark World 2023 organizers, as some of the discussions took place during LDW 2023.}

\providecommand{\href}[2]{#2}\begingroup\raggedright\endgroup

\end{document}